\documentclass[submission,copyright,creativecommons]{eptcs}
\providecommand{\event}{INFINITY 2012} % Name of the event you are submitting to

\title{Petri Nets with Time and Cost\\
(Tutorial)}
\author{Parosh Aziz Abdulla
\institute{Department of Information Technology\\ 
Uppsala University\thanks{This work is supported by {\sc Upmarc},
The Uppsala Programming for Multicore Architectures Research Center.}\\
Sweden}
\email{parosh@it.uu.se}
\and
Richard Mayr
\institute{School of Informatics, LFCS\\
University of Edinburgh\\
United Kingdom}
\email{homepages.inf.ed.ac.uk/rmayr/}
}
\def\titlerunning{Timed Priced  Petri Nets}
\def\authorrunning{P. A. Abdulla, R. Mayr}

\usepackage{pgf}
\usepackage{tikz}
\usepackage{clrscode}
\usepackage{amsmath}
\usepackage{amsfonts}
\usepackage{scalefnt}

\usetikzlibrary{automata}
\usetikzlibrary{petri}
\usetikzlibrary{shadows}
\usetikzlibrary{positioning}
\usetikzlibrary{decorations.pathmorphing}
\usetikzlibrary{decorations.shapes}
\usetikzlibrary{shapes}
\usetikzlibrary{backgrounds}

\usetikzlibrary{fit}
\usetikzlibrary{arrows}
\usetikzlibrary{calc}
\usetikzlibrary{mindmap}
\usetikzlibrary{matrix}
\usepackage{pgffor}
\usepackage{xcolor}

\tikzset{para/.style={shade,bottom color=#1,shading angle=180}}
\tikzset{link/.style={-stealth,line width=2pt}}
\tikzset{zone/.style={circle,shading=radial,inner color=#1!yellow,draw=blue!50,line width=2pt}}

\tikzstyle{moveprice}=[fill=black!20,inner sep=2pt,rounded corners]

\tikzstyle{west-node}=[anchor=west]

\tikzstyle{regionnode}=[fill=black,rounded corners,inner sep=1mm]
\tikzstyle{Regionnode}=[fill=gray,draw=black,rounded corners,inner sep=1mm]
\tikzstyle{moveplace}=[draw,inner sep=0.8pt,text=black,circle,minimum size=2.5mm]
\tikzstyle{large-moveplace}=[draw,inner sep=0pt,text=black,circle,minimum size=5mm,west-node]

\tikzstyle{placeprice}=[fill=black!20,inner sep=1pt,circle]
\tikzstyle{intervalnode}=[fill=yellow!20,inner sep=1pt]

\tikzstyle{single-moveplace-bg}=[rounded corners,fill=black,rectangle,minimum width = 7mm,minimum height=7mm,west-node]
\tikzstyle{double-moveplace-bg}=[west-node,rounded corners,fill=black,rectangle,minimum width = 13mm,minimum height=7mm]
\tikzstyle{triple-moveplace-bg}=[west-node,rounded corners,fill=black,rectangle,minimum width = 19mm,minimum height=7mm]

\tikzstyle{marking-place}=[draw,inner sep=0pt,circle,minimum size=4mm,west-node]
\tikzstyle{red-marking-place}=[marking-place,fill=red!80!black,text=white]
\tikzstyle{blue-marking-place}=[marking-place,fill=blue!90!black,text=white]
\tikzstyle{green-marking-place}=[marking-place,fill=green,text=black]
\tikzstyle{orange-marking-place}=[marking-place,fill=orange,text=black]
\tikzstyle{white-marking-place}=[marking-place,fill=white,text=black]
\tikzstyle{black-marking-place}=[marking-place,fill=black,text=black]

\tikzstyle{single-markingplace-bg}=[rounded corners,fill=black,rectangle,minimum width = 6mm,minimum height=6mm,west-node]
\tikzstyle{double-markingplace-bg}=[west-node,rounded corners,fill=black,rectangle,minimum width = 11mm,minimum height=7mm]
\tikzstyle{triple-markingplace-bg}=[west-node,rounded corners,fill=black,rectangle,minimum width = 16mm,minimum height=7mm]

\tikzstyle{ball}=[draw,circle,minimum size=1mm,color=black]
\tikzstyle{red-ball}=[ball,fill=red!80!black]
\tikzstyle{blue-ball}=[ball,fill=blue!90!black]
\tikzstyle{white-ball}=[ball,fill=white]
\tikzstyle{green-ball}=[ball,fill=green]
\tikzstyle{orange-ball}=[ball,fill=orange]

\tikzstyle{regionnode}=[fill=black,rounded corners,inner sep=1mm,font=\footnotesize]
\tikzstyle{Regionnode}=[fill=gray,draw=black,rounded corners,inner sep=1mm,font=\footnotesize]
\tikzstyle{move-place}=[anchor=west,draw,inner sep=2pt,circle,minimum size=5mm,font=\tiny]
\tikzstyle{red-move-place}=  [move-place, fill=red!80!black ,text=white]
\tikzstyle{blue-move-place}= [move-place, fill=blue!90!black,text=white]
\tikzstyle{green-move-place}=[move-place, fill=green,text=black]
\tikzstyle{orange-move-place}=[move-place, fill=orange,text=black]
\tikzstyle{white-move-place}=[move-place, fill=white,text=black]
\tikzstyle{black-move-place}=[move-place, fill=black,text=black]
\tikzstyle{yellow-move-place}=[move-place, color=yellow,text=yellow]

\tikzstyle{my-place}=[place,minimum size = 7mm,inner sep=0mm,draw,drop shadow]

\tikzstyle{red-place}=[my-place,fill=red!80!black,text=white]
\tikzstyle{blue-place}=[my-place,fill=blue!90!black,text=white]
\tikzstyle{green-place}=[my-place,fill=green,text=black]
\tikzstyle{orange-place}=[my-place,fill=orange,text=black]
\tikzstyle{white-place}=[my-place,fill=white,text=black]

\tikzstyle{black-bg}=[rounded corners,fill=black]
\tikzstyle{yellow-bg}=[rounded corners,fill=yellow]
\tikzstyle{arrow-node}=[anchor=west,font=\Large]
\tikzstyle{moveprice}=[fill=black!20,inner sep=2pt,rounded corners,font=\footnotesize]
\tikzstyle{movelabel}=[inner sep=2pt,rounded corners,font=\footnotesize]

\pgfdeclarelayer{background}
\pgfdeclarelayer{foreground}
\pgfsetlayers{background,main,foreground}

\newcommand{\hide}[1]{}

\newcommand{\set}[1]{\left\{#1\right\}}

\newcommand{\tuple}[1]{\left(#1\right)}
\newcommand{\setcomp}[2]{\left\{#1|\; #2\right\}}

\newcommand{\denotationof}[1]{[\![#1]\!]}

\newcommand{\nat}{{\mathbb N}}

\newcommand{\nnreals}{{\mathbb R}_{\geq 0}}

\newcommand{\conf}{c}

\newcommand{\comp}{\pi}

\newcommand{\Intervals}{{\it Intrv}}

\newcommand{\oointrvl}[2]{({#1}:{#2})}

\newcommand{\ptpn}{{\cal N}}
\newcommand{\ptpntuple}{\tuple{\places,\transitions,\costfun}}

\newcommand{\places}{P}

\newcommand{\place}{p}
\newcommand{\initp}{\place_{\it init}}
\newcommand{\finalp}{\place_{\it fin}}

\newcommand{\transitions}{T}
\newcommand{\transition}{t}

\newcommand{\inputs}{\it In}

\newcommand{\outputs}{\it Out}

\newcommand{\costfun}{{\it Cost}}

\newcommand{\costof}[1]{\costfun\left(#1\right)}

\newcommand{\marking}{M}
\newcommand{\markings}{{\tt M}}

\newcommand{\movesto}[1]{\stackrel{#1}{\longrightarrow}}
\newcommand{\amovesto}[1]{{\stackrel{#1}{\longrightarrow}}_A}
\newcommand{\bmovesto}[1]{{\stackrel{#1}{\longrightarrow}}_B}

\newcommand{\disc}{{\it Disc}}
\newcommand{\discmovesto}{\longrightarrow_\disc}

\newcommand{\pre}{\textit{Pre}}

\newcommand{\region}{r}
\newcommand{\finalregion}{\region_{\it fin}}
\newcommand{\initregion}{\region_{\it init}}

\newcommand{\maxval}{{\it cmax}}

\newcommand{\initmarking}{\marking_{\it init}}

\newcommand{\w}{{\it w}}
\newcommand{\z}{{\it z}}

\newcommand{\ucof}[1]{{#1}\!\uparrow}
\newcommand{\ordering}{\sqsubseteq}

\newcommand{\threshold}{v}

\newcommand{\white}[1]{\textcolor{white}{#1}}
\newcommand{\remainder}{u}
\newcommand{\fordering}{\sqsubseteq_{\it free}}
\newcommand{\allordering}{\sqsubseteq_{\it all}}
\newcommand{\confset}{S}
\newcommand{\costconfset}{C}
\newcommand{\confs}{{\tt S}}
\newcommand{\ts}{{\mathcal T}}
\newcommand{\freeucof}[1]{{#1}\!\uparrow\!{\it free}}
\newcommand{\allucof}[1]{{#1}\!\uparrow\!{\it all}}
\newcommand{\allminof}[1]{{\it min}_{\it all}\left(#1\right)}
\newcommand{\freeminof}[1]{{\it min}_{\it free}\left(#1\right)}

\begin{document}
\maketitle

\section{Introduction}
Petri nets \cite{Petri:thesis,Peterson:survey}
are a widely used model for the study and analysis of concurrent systems.
Many different formalisms have been proposed which extend 
Petri nets with clocks and real-time constraints, leading to 
various definitions of {\it Timed Petri nets ({\sc Tpn}s)}
(see \cite{Bowden:TPN:Survey,BCHLR-atva2005} for surveys).

In parallel, there have been several works on extending the
model of timed automata \cite{AD:timedautomata} with {\it prices}
({\it weights}) (see e.g., \cite{AlurTP01:WTA,Kim:etal:priced,BouyerBBR:PTA}).
Weighted timed automata are suitable models for embedded systems, where 
we have to take into consideration the fact that the behavior of the system
may be constrained by the consumption of different types of resources.
Concretely, weighted timed automata
extend classical timed automata 
with a cost function $\costfun$ 
that maps every location and every transition to a 
nonnegative integer (or rational) number.
For a transition, $\costfun$ gives the cost of 
performing the transition. 
For a location, $\costfun$ gives the cost per time unit for 
staying in the location. 
In this manner, we can define, for each computation of the system, the accumulated 
cost of staying in locations and performing transitions 
along the computation.

In this tutorial, we recall, through a sequence of examples,
a very expressive model, introduced in \cite{abdulla2011computing},
 that subsumes the above models. 
{\em Priced Timed Petri Nets} ({\sc Ptpn})
are a generalization of classic Petri nets \cite{Petri:thesis} with
real-valued (i.e., continuous-time) clocks, real-time constraints, 
and prices for computations.

In a {\sc Ptpn}, each token is equipped with a real-valued clock, representing the age of the token.
The firing conditions of a transition include the usual ones for Petri nets.
Additionally, each arc between a place and a transition is labeled with a 
time-interval whose bounds are natural numbers (or possibly $\infty$ as upper
bound). These intervals can be open, closed or half
open. Like in timed automata, this is used to encode strict or non-strict
inequalities that describe constraints on the real-valued clocks.
When firing a transition, tokens which 
are removed from or added to places must have ages lying in the 
intervals of the corresponding transition arcs.

We assign a cost to computations via a cost function $\costfun$ that
maps transitions and places of the Petri net to natural numbers.
For a transition $t$, $\costfun(t)$ gives the cost of
performing the transition, while for a place $p$,
$\costfun(p)$ gives the cost per time unit per token in the place.
The total cost of a computation is given by the sum of all costs of
fired transitions plus the storage costs for storing certain numbers of tokens
in certain places for certain times during the computation.
Like in priced timed automata, having integers as costs and time bounds is not a restriction, because
the case of rational numbers can be reduced to the integer case.

It should be noted that {\sc Ptpn} are infinite-state in several different ways.
First, the Petri net itself is unbounded. So the number of tokens (and thus
the number of clocks) can grow beyond any bound, i.e., the {\sc Ptpn} can create and
destroy arbitrarily many clocks (unlike timed automata). 
Secondly, every single clock value is a real number of which there are 
uncountably many.

In \cite{abdulla2011computing}
we study the cost to reach a given control-state 
in a {\sc Ptpn}. In Petri net terminology, this is called a control-state
reachability problem or a coverability problem. 
The related reachability problem (i.e., reaching a particular
configuration) is undecidable for both continuous-time and discrete-time
TPN \cite{PCVC99:TPN:Nondecidability}, even without taking costs into
account.
Our goal is to compute the optimal
cost for moving to a control state (equivalently for covering a set of markings).
In general, a cost-optimal computation may not exist
(e.g., even in priced timed automata it can happen that there is no computation
of cost $0$, but there exist computations of cost $\le \epsilon$ for every 
$\epsilon > 0$).
We show that the {\em infimum} of the costs to reach a given control-state 
is computable, provided that all transition and place costs are non-negative.

\paragraph{Outline.}
In the next section we introduce {\sc Ptpn}s.
In Section~\ref{delta:section} we describe a special type of computations
that are sufficient to solve the cost-optimality problem.
We introduce a symbolic encoding of infinite sets of markings in 
Section~\ref{regions:section}, and describe a symbolic
algorithm for solving the cost-optimality problem in
Section~\ref{solution:section}.
Finally, in Section~\ref{conclusions:section}, we 
give conclusions and directions for future work.

\section{Timed Petri Nets}
\label{ptpn:section}

In this section, we introduce Priced Timed Petri Nets, 
the set of markings, the transition relation it induces, and
the coverability problem.

We use $\nat$ and $\nnreals$ to denote the sets
of natural numbers (including 0) and nonnegative reals respectively. 
We use a set $\Intervals$ of intervals. An open interval is 
written as $\oointrvl  w z$ 
 where $\w\in\nat$ and $\z\in\nat\cup\set{\infty}$. Intervals can also 
be closed in one or both directions, 
 e.g. $[w:z]$ is closed in both directions  and
  $[w:z)$ is closed to the left and open to the right.

\paragraph{Model.}
A {\em Priced Timed Petri Net ({\sc Ptpn})} is a tuple $\ptpn=\ptpntuple$ where
$\places$ is a finite set of places.
$\transitions$ is a finite set of transitions,
where each transition $\transition \in \transitions$ is of the form
$\transition = (\inputs,\outputs)$.
We have that 
$\inputs$ and $\outputs$
are finite multisets over $\places \times \Intervals$ which define the
input-arcs and output-arcs of $t$, respectively.
$\costfun:\places \cup \transitions \to \nat$ 
is the cost function assigning
firing costs to transitions and storage costs to places.
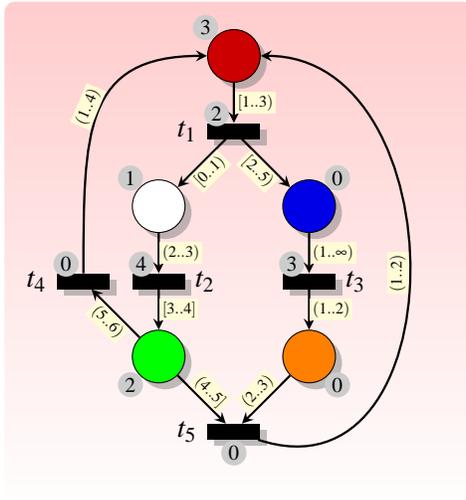
\begin{figure}
\center
\begin{tikzpicture}[
background rectangle/.style=
{top color=red!20,rounded corners}, show background rectangle,
>=stealth,every transition/.style={minimum width = 7mm,minimum height=2mm,inner sep=0mm,drop shadow,draw=none,fill=black}]

%%%%%%%%%%%%%%%%%%%  p1   %%%%%%%%%%%%%%%

  \node at (0,0) [red-place,label={[placeprice]135:{\scriptsize $3$}}] (p1){};

%%%%%%%%%%%%%%%%%%%  p2   %%%%%%%%%%%%%%%

  \node at (-10mm,-20mm)[white-place,label={[placeprice]135:{\scriptsize $1$}}] (p2){};

%%%%%%%%%%%%%%%%%%%  p3   %%%%%%%%%%%%%%%
     
  \node at (10mm,-20mm)[blue-place,label={[placeprice]45:{\scriptsize $0$}}] (p3){};

%%%%%%%%%%%%%%%%%%%  p4   %%%%%%%%%%%%%%%

  \node at (-10mm,-40mm)[green-place,label={[placeprice]225:{\scriptsize $2$}}] (p4){};

%%%%%%%%%%%%%%%%%%%  p5  %%%%%%%%%%%%%%%

     \node at (10mm,-40mm)[orange-place,label={[placeprice]315:{\scriptsize $0$}}] (p5){};

%%%%%%%%%%%%%%%%%%%  t1   %%%%%%%%%%%%%%%

     \node at (0mm, -10mm) [transition,label={180:$t_1$},label={[placeprice]135:{\scriptsize $2$}}] (t1){};

%%%%%%%%%%%%%%%%%%%  t2   %%%%%%%%%%%%%%%

     \node at (-10mm, -30mm) [transition,draw=none,fill=black,label={0:$t_2$},label={[placeprice]135:{\scriptsize $4$}}] (t2){};

%%%%%%%%%%%%%%%%%%%  t3    %%%%%%%%%%%%%%%

     \node at (10mm, -30mm) [transition,fill=black,label={0:$t_3$},label={[placeprice]135:{\scriptsize $3$}}] (t3){};

%%%%%%%%%%%%%%%%%%%  t4   %%%%%%%%%%%%%%%
     
     \node at (-20mm, -30mm) [transition,label={180:$t_4$},label={[placeprice]135:{\scriptsize $0$}}] (t4){};

%%%%%%%%%%%%%%%%%%%  t5   %%%%%%%%%%%%%%%

     \node at (0mm, -50mm) [transition,label={180:$t_5$},label={[placeprice]270:{\scriptsize $0$}}] (t5){};

%%%%%%%%%%%%%%%%%%%  Edges  %%%%%%%%%%%%%%%%%%%%

   \draw [->,line width=0.3mm] (p1) -- node[right,intervalnode]{\tiny $[1..3)$} (t1);
   \draw [->,line width=0.3mm] (t1) -- node[below,intervalnode,sloped]{\tiny $[0..1)$}  (p2);
   \draw [->,line width=0.3mm] (t1) -- node[below,intervalnode,sloped]{\tiny $[2..5)$}   (p3);
   \draw [->,line width=0.3mm] (p2) -- node[right,intervalnode]{\tiny $(2..3)$} (t2);
   \draw [->,line width=0.3mm] (p3) -- node[right,intervalnode]{\tiny $(1..\infty)$} (t3);
   \draw [->,line width=0.3mm] (t2) -- node[right,intervalnode]{\tiny $[3..4]$} (p4);
   \draw [->,line width=0.3mm] (t3) -- node[right,intervalnode]{\tiny $(1..2)$} (p5);
   \draw [->,line width=0.3mm] (p4) -- node[above,intervalnode,sloped]{\tiny $(4..5]$} (t5);
   \draw [->,line width=0.3mm] (p5) -- node[above,intervalnode,sloped]{\tiny $(2..3)$} (t5);
   \draw [->,line width=0.3mm,out=90,in=180] (t4) .. controls  ($(t4)+(0mm,20mm)$) and 
    ($(p1)+(-20mm,0)$) .. node[above,intervalnode,sloped]{\tiny $(1..4)$}  (p1);
   \draw [->,line width=0.3mm] (p4) -- node[below,intervalnode,sloped]{\tiny $(5..6)$} (t4);
   \draw  [->,line width=0.3mm] (t5) .. controls  ($(t5)+(30mm,-10mm)$) and 
    ($(p1)+(30mm,0)$) .. node[above,intervalnode,sloped]{\tiny $(1..2)$} (p1);

\end{tikzpicture}
\caption{A Price Timed Petri Net.}
\label{ptpn:figure}
\end{figure}
Figure~\ref{ptpn:figure} shows an example of a {\sc Ptpn}
with five places:
\tikz{\node[red-ball]{};},
\tikz{\node[white-ball]{};},
\tikz{\node[blue-ball]{};},
\tikz{\node[green-ball]{};},
\tikz{\node[orange-ball]{};}, and
five transitions:
$t_1,t_2,t_3,t_4,t_5$.
The transition $t_1$ has an input arc from 
\tikz{\node[red-ball]{};} 
labeled with the interval 
$[1..3]$, and
two output arcs to
\tikz{\node[white-ball]{};} 
and
\tikz{\node[blue-ball]{};},
 labeled with
the intervals $(0..1)$ and
$[2..5]$ respectively.
The price (cost) associated with 
\tikz{\node[red-ball]{};},
is $3$, while the price associated with $t_1$ is $2$.
We let $\maxval$ denote the maximum integer appearing on the arcs of a given
{\sc Ptpn}.
In Figure~\ref{ptpn:figure}, we have $\maxval=6$.

\begin{figure}
\begin{tikzpicture}[
>=stealth,every transition/.style={minimum width = 7mm,minimum height=2mm,inner sep=0mm,drop shadow,draw=none,fill=black}]

\node[rectangle,rounded corners,anchor=north,top color=red!20,minimum width=51mm,minimum height=60mm] at (-1mm,6mm){};
%%%%%%%%%%%%%%%%%%%  p1   %%%%%%%%%%%%%%%

  \node at (0,0) [red-place,label={[placeprice]135:{\scriptsize $3$}}] (p1){};

    \node at (p1) [] {\tiny
      \begin{tabular}{l}
        \white{$7.93$}\\
        \white{$1.08$}
      \end{tabular}
    };

%%%%%%%%%%%%%%%%%%%  p2   %%%%%%%%%%%%%%%

  \node at (-10mm,-20mm)[white-place,label={[placeprice]135:{\scriptsize $1$}}] (p2){};

    \node at (p2) [] {\tiny
      \begin{tabular}{l}
        $2.32$
      \end{tabular}
    };

%%%%%%%%%%%%%%%%%%%  p3   %%%%%%%%%%%%%%%
     
  \node at (10mm,-20mm)[blue-place,label={[placeprice]45:{\scriptsize $0$}}] (p3){};

    \node at (p3) [] {\tiny
      \begin{tabular}{l}
        \white{$2.11$}
      \end{tabular}
    };

%%%%%%%%%%%%%%%%%%%  p4   %%%%%%%%%%%%%%%

  \node at (-10mm,-40mm)[green-place,label={[placeprice]225:{\scriptsize $2$}}] (p4){};

    \node at (p4) [] {\tiny
      \begin{tabular}{l}
        $0.25$\\
        $8.36$
      \end{tabular}
    };

%%%%%%%%%%%%%%%%%%%  p5  %%%%%%%%%%%%%%%

  \node at (10mm,-40mm)[orange-place,label={[placeprice]315:{\scriptsize $0$}}] (p5){};

    \node at (p5) [] {\tiny
      \begin{tabular}{l}
        $4.21$
      \end{tabular}
    };

%%%%%%%%%%%%%%%%%%%  t1   %%%%%%%%%%%%%%%

     \node at (0mm, -10mm) [transition,label={180:$t_1$},label={[placeprice]135:{\scriptsize $2$}}] (t1){};

%%%%%%%%%%%%%%%%%%%  t2   %%%%%%%%%%%%%%%

     \node at (-10mm, -30mm) [transition,draw=none,fill=black,label={0:$t_2$},label={[placeprice]135:{\scriptsize $4$}}] (t2){};

%%%%%%%%%%%%%%%%%%%  t3    %%%%%%%%%%%%%%%

     \node at (10mm, -30mm) [transition,fill=black,label={0:$t_3$},label={[placeprice]135:{\scriptsize $3$}}] (t3){};

%%%%%%%%%%%%%%%%%%%  t4   %%%%%%%%%%%%%%%
     
     \node at (-20mm, -30mm) [transition,label={180:$t_4$},label={[placeprice]135:{\scriptsize $0$}}] (t4){};

%%%%%%%%%%%%%%%%%%%  t5   %%%%%%%%%%%%%%%

     \node at (0mm, -50mm) [transition,label={180:$t_5$},label={[placeprice]270:{\scriptsize $0$}}] (t5){};

%%%%%%%%%%%%%%%%%%%  Edges  %%%%%%%%%%%%%%%%%%%%

   \draw [->,line width=0.3mm] (p1) -- node[right,intervalnode]{\tiny $[1..3)$} (t1);
   \draw [->,line width=0.3mm] (t1) -- node[below,intervalnode,sloped]{\tiny $[0..1)$}  (p2);
   \draw [->,line width=0.3mm] (t1) -- node[below,intervalnode,sloped]{\tiny $[2..5)$}   (p3);
   \draw [->,line width=0.3mm] (p2) -- node[right,intervalnode]{\tiny $(2..3)$} (t2);
   \draw [->,line width=0.3mm] (p3) -- node[right,intervalnode]{\tiny $(1..\infty)$} (t3);
   \draw [->,line width=0.3mm] (t2) -- node[right,intervalnode]{\tiny $[3..4]$} (p4);
   \draw [->,line width=0.3mm] (t3) -- node[right,intervalnode]{\tiny $(1..2)$} (p5);
   \draw [->,line width=0.3mm] (p4) -- node[above,intervalnode,sloped]{\tiny $(4..5]$} (t5);
   \draw [->,line width=0.3mm] (p5) -- node[above,intervalnode,sloped]{\tiny $(2..3)$} (t5);
   \draw [->,line width=0.3mm,out=90,in=180] (t4) .. controls  ($(t4)+(0mm,20mm)$) and 
    ($(p1)+(-20mm,0)$) .. node[above,intervalnode,sloped]{\tiny $(1..4)$}  (p1);
   \draw [->,line width=0.3mm] (p4) -- node[below,intervalnode,sloped]{\tiny $(5..6)$} (t4);
   \draw  [->,line width=0.3mm] (t5) .. controls  ($(t5)+(30mm,-10mm)$) and 
    ($(p1)+(30mm,0)$) .. node[above,intervalnode,sloped]{\tiny $(1..2)$} (p1);

\begin{scope}[xshift=4cm,yshift=-2cm]

  \node [red-move-place,name=n1] {$7.93$};
  \node at ($(n1.east)+(1mm,0mm)$) [red-move-place,name=n2] {$1.08$};
  \node at ($(n2.east)+(1mm,0mm)$)  [white-move-place,name=n3] {$2.32$};
  \node at ($(n3.east)+(1mm,0mm)$)  [blue-move-place,name=n4] {$2.11$};
  \node at ($(n4.east)+(1mm,0mm)$)  [green-move-place,name=n5] {$0.25$};
  \node at ($(n5.east)+(1mm,0mm)$)  [green-move-place,name=n6] {$8.36$};
  \node at ($(n6.east)+(1mm,0mm)$)  [orange-move-place,name=n7] {$4.21$};

\begin{pgfonlayer}{background}
\node[black-bg,fit=(n1) (n2) (n3) (n4) (n5) (n6) (n7)]{};
\end{pgfonlayer}

\end{scope}

\end{tikzpicture}
\label{marking:fig}
\caption{A marking $\marking$ and its representation.}
\end{figure}
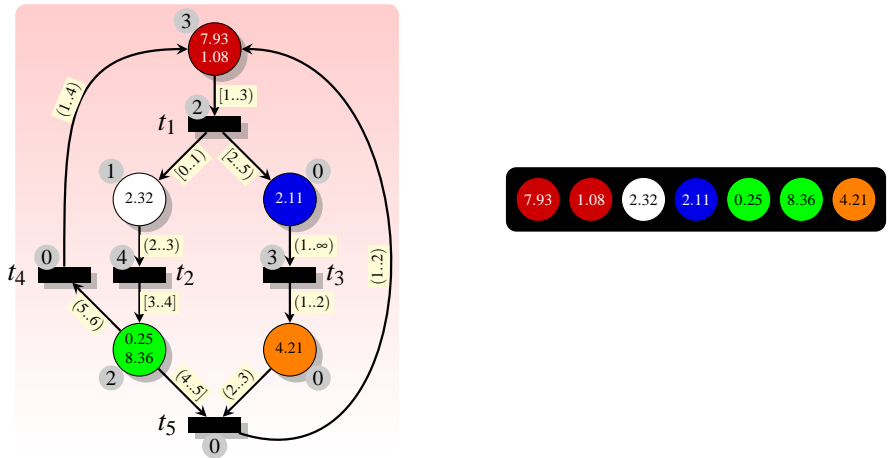

\paragraph{Markings.}
A marking is
a multiset over $\places\times\nnreals$.
The marking $\marking$ defines the numbers  and ages
of tokens in each place in the net.
In Figure~\ref{marking:fig}, we show an example of a marking
$\marking$.
The marking assigns two tokens in 
\tikz{\node[red-ball]{};}, with ages
$7.93$ and $1.08$, respectively.
We will represent markings by lists of ``colored balls''
with real numbers inside.
Each ball represents one token in the marking.
The color describes the place in which the token resides, 
while the number represents the age of the token
(see Figure~\ref{marking:fig}).

\begin{figure}
\center
\begin{tikzpicture}[]

  \node[name=dummy]   {};
  
  \node[red-move-place,name=p1] at (dummy)  {$0.0$};
  \begin{pgfonlayer}{background}
    \node[black-bg,fit=(p1) ]{};
  \end{pgfonlayer}

  \node[arrow-node,name=arrow] at ($(p1.east)+(2mm,0mm)$) {$\longrightarrow$};
  \node[movelabel] at ($(arrow.center)+(0mm,3.5mm)$) {$1.7$};
  \node[moveprice] at ($(arrow.center)+(0mm,-3.5mm)$) {$5.1$};

  \node[red-move-place,name=p1] at ($(arrow.east)+(2mm,0)$) {$1.7$};
  \begin{pgfonlayer}{background}
    \node[black-bg,fit=(p1) ]{};
  \end{pgfonlayer}

  \node[arrow-node,name=arrow] at ($(p1.east)+(2mm,0mm)$) {$\longrightarrow$};
  \node[movelabel] at ($(arrow.center)+(0mm,3.5mm)$) {$t_1$};
  \node[moveprice] at ($(arrow.center)+(0mm,-3.5mm)$) {$2.0$};

  \node[white-move-place,name=p1] at ($(arrow.east)+(2mm,0)$) {$0.1$};
  \node[blue-move-place,name=p2] at ($(p1.east)+(1mm,0)$) {$3.1$};
  \begin{pgfonlayer}{background}
    \node[black-bg,fit=(p1) (p2)]{};
  \end{pgfonlayer}

  \node[arrow-node,name=arrow] at ($(p2.east)+(2mm,0mm)$) {$\longrightarrow$};
  \node[movelabel] at ($(arrow.center)+(0mm,3.5mm)$) {$2.3$};
  \node[moveprice] at ($(arrow.center)+(0mm,-3.5mm)$) {$2.3$};

  \node[white-move-place,name=p1] at ($(arrow.east)+(2mm,0)$) {$2.4$};
  \node[blue-move-place,name=p2] at ($(p1.east)+(1mm,0)$) {$5.4$};
  \begin{pgfonlayer}{background}
    \node[black-bg,fit=(p1) (p2)]{};
  \end{pgfonlayer}

  \node[arrow-node,name=arrow] at ($(p2.east)+(2mm,0mm)$) {$\longrightarrow$};
  \node[movelabel] at ($(arrow.center)+(0mm,3.5mm)$) {$t_2$};
  \node[moveprice] at ($(arrow.center)+(0mm,-3.5mm)$) {$4$};

  \node[green-move-place,name=p1] at ($(arrow.east)+(2mm,0)$) {$3.6$};
  \node[blue-move-place,name=p2] at ($(p1.east)+(1mm,0)$) {$5.4$};
  \begin{pgfonlayer}{background}
    \node[black-bg,fit=(p1) (p2)]{};
  \end{pgfonlayer}

  \node[arrow-node,name=arrow] at ($(p2.east)+(2mm,0mm)$) {$\longrightarrow$};
  \node[movelabel] at ($(arrow.center)+(0mm,3.5mm)$) {$1.5$};
  \node[moveprice] at ($(arrow.center)+(0mm,-3.5mm)$) {$3$};

  \node[name=dummy] at ($(dummy)+(0mm,-20mm)$)  {};
  \node[green-move-place,name=p1] at (dummy) {$5.1$};
  \node[blue-move-place,name=p2] at ($(p1.east)+(1mm,0)$) {$6.9$};
  \begin{pgfonlayer}{background}
    \node[black-bg,fit=(p1) (p2)]{};
  \end{pgfonlayer}

  \node[arrow-node,name=arrow] at ($(p2.east)+(2mm,0mm)$) {$\longrightarrow$};
  \node[movelabel] at ($(arrow.center)+(0mm,3.5mm)$) {$t_4$};
  \node[moveprice] at ($(arrow.center)+(0mm,-3.5mm)$) {$0$};

  \node[red-move-place,name=p1] at ($(arrow.east)+(2mm,0)$) {$2.0$};
  \node[blue-move-place,name=p2] at ($(p1.east)+(1mm,0)$) {$6.9$};
  \begin{pgfonlayer}{background}
    \node[black-bg,fit=(p1) (p2)]{};
  \end{pgfonlayer}

  \node[arrow-node,name=arrow] at ($(p2.east)+(2mm,0mm)$) {$\longrightarrow$};
  \node[movelabel] at ($(arrow.center)+(0mm,3.5mm)$) {$t_1$};
  \node[moveprice] at ($(arrow.center)+(0mm,-3.5mm)$) {$2$};

  \node[white-move-place,name=p1] at ($(arrow.east)+(2mm,0)$) {$0.8$};
  \node[blue-move-place,name=p2] at ($(p1.east)+(2mm,0)$) {$6.9$};
  \node[blue-move-place,name=p3] at ($(p2.east)+(1mm,0)$) {$3.1$};
  \begin{pgfonlayer}{background}
    \node[black-bg,fit=(p1) (p2) (p3)]{};
  \end{pgfonlayer}

  \node[arrow-node,name=arrow] at ($(p3.east)+(2mm,0mm)$) {$\longrightarrow$};
  \node[movelabel] at ($(arrow.center)+(0mm,3.5mm)$) {$1.5$};
  \node[moveprice] at ($(arrow.center)+(0mm,-3.5mm)$) {$1.5$};

  \node[white-move-place,name=p1] at ($(arrow.east)+(2mm,0)$) {$2.3$};
  \node[blue-move-place,name=p2] at ($(p1.east)+(2mm,0)$) {$8.4$};
  \node[blue-move-place,name=p3] at ($(p2.east)+(1mm,0)$) {$4.6$};
  \begin{pgfonlayer}{background}
    \node[black-bg,fit=(p1) (p2) (p3)]{};
  \end{pgfonlayer}

  \node[arrow-node,name=arrow] at ($(p3.east)+(2mm,0mm)$) {$\longrightarrow$};
  \node[movelabel] at ($(arrow.center)+(0mm,3.5mm)$) {$t_2$};
  \node[moveprice] at ($(arrow.center)+(0mm,-3.5mm)$) {$4$};

  \node[name=dummy] at ($(dummy)+(0mm,-20mm)$)  {};
  \node[green-move-place,name=p1] at (dummy) {$3.7$};
  \node[blue-move-place,name=p2] at ($(p1.east)+(1mm,0)$) {$8.4$};
  \node[blue-move-place,name=p3] at ($(p2.east)+(1mm,0)$) {$4.6$};
  \begin{pgfonlayer}{background}
    \node[black-bg,fit=(p1) (p2) (p3)]{};
  \end{pgfonlayer}

  \node[arrow-node,name=arrow] at ($(p3.east)+(2mm,0mm)$) {$\longrightarrow$};
  \node[movelabel] at ($(arrow.center)+(0mm,3.5mm)$) {$t_3$};
  \node[moveprice] at ($(arrow.center)+(0mm,-3.5mm)$) {$3$};

  \node[green-move-place,name=p1] at ($(arrow.east)+(2mm,0)$) {$3.7$};
  \node[orange-move-place,name=p2] at ($(p1.east)+(2mm,0)$) {$1.1$};
  \node[blue-move-place,name=p3] at ($(p2.east)+(1mm,0)$) {$4.6$};
  \begin{pgfonlayer}{background}
    \node[black-bg,fit=(p1) (p2) (p3)]{};
  \end{pgfonlayer}

  \node[arrow-node,name=arrow] at ($(p3.east)+(2mm,0mm)$) {$\longrightarrow$};
  \node[movelabel] at ($(arrow.center)+(0mm,3.5mm)$) {$1$};
  \node[moveprice] at ($(arrow.center)+(0mm,-3.5mm)$) {$2.0$};

  \node[green-move-place,name=p1] at ($(arrow.east)+(2mm,0)$) {$4.7$};
  \node[orange-move-place,name=p2] at ($(p1.east)+(2mm,0)$) {$2.1$};
  \node[blue-move-place,name=p3] at ($(p2.east)+(1mm,0)$) {$5.6$};
  \begin{pgfonlayer}{background}
    \node[black-bg,fit=(p1) (p2) (p3)]{};
  \end{pgfonlayer}

  \node[arrow-node,name=arrow] at ($(p3.east)+(2mm,0mm)$) {$\longrightarrow$};
  \node[movelabel] at ($(arrow.center)+(0mm,3.5mm)$) {$t_5$};
  \node[moveprice] at ($(arrow.center)+(0mm,-3.5mm)$) {$0$};

  \node[red-move-place,name=p1] at ($(arrow.east)+(2mm,0)$) {$1.5$};
  \node[blue-move-place,name=p2] at ($(p1.east)+(1mm,0)$) {$5.6$};
  \begin{pgfonlayer}{background}
    \node[black-bg,fit=(p1) (p2)]{};
  \end{pgfonlayer}

\end{tikzpicture}
\label{comp:fig}
\caption{A computation $\comp$. Above each $\longrightarrow$
in the computation we show the transition that has fired, and 
below each step we show the cost of the step.}
\end{figure}
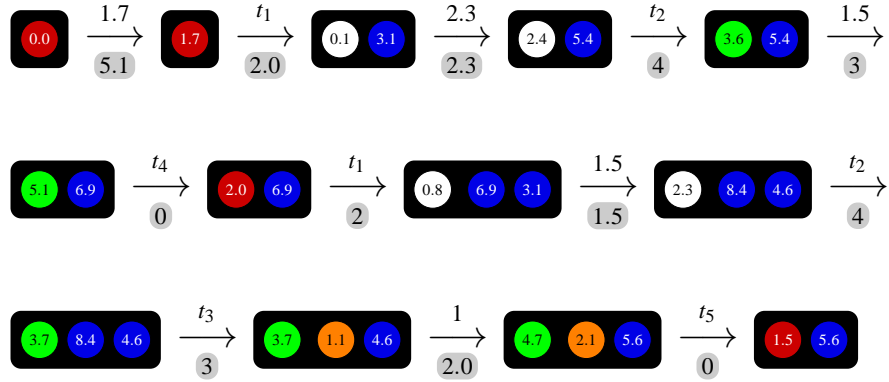

\paragraph{Computations.}
We define two transition relations on the 
set of configurations: timed transition
and discrete transition.
 A {\em timed transition} increases the age of each token
by the same real number.
A {\it discrete transition} represents the effect of {\it firing}
a transition $\transition$ in the {\sc Ptpn}.
More precisely, for each input arc to the transition, we remove
a token from the corresponding input place, whose age
lies in the relevant interval.
Also, for each input arc to the transition, we add a new
token to the corresponding place.
The age of the newly generated token is chosen non-deterministically
from the relevant interval.
Performing a discrete transition implies paying a cost
which is equal to the cost of the transition.
When performing a timed transition, we pay a cost per each token and time
unit
that is equal to the cost of the place in which the token resides.
A {\it computation} is a sequence of discrete and timed transitions.
The cost of  a computation is the accumulated cost of all the 
transitions in the computation.
Figure~\ref{comp:fig} shows an example of a computation $\comp$.
It starts from an initial marking where we have a single token in
\tikz{\node[red-ball]{};} with age $0$.
In the seventh step of $\comp$, transition $t_1$ fires removing one token
from \tikz{\node[red-ball]{};} with age $2$.
The age belongs to the interval $[1..3)$ 
(which is the interval on the arc from 
\tikz{\node[red-ball]{};} to $t_1$).
At the same time, it adds two new tokens with ages 
$0.8$ and $3.1$ to the places
\tikz{\node[white-ball]{};} resp.\ \tikz{\node[blue-ball]{};}.
The cost of this step is equal to $2$.
The eighth step is a timed transition of length $1.5$,
where the ages of all tokens are increased by $1.5$.
The cost of the step is determined by the number of tokens
in each place and the cost of the place, i.e.,
$1.5 \times (1 \times 1 + 2 \times 0)=1.5$ (the cost of
\tikz{\node[white-ball]{};} and \tikz{\node[blue-ball]{};}
are $1$ resp.\ $0$).
The total cost of $\comp$ is given by
$\costof\pi =
        5.1+2+2.3+
        4+3+0+2+
        1.5+4+3+
        2+0=28.9$.

For a place $\place$, we define $\marking_\place$
to be the set of markings which put at least one token
in the place $\place$ (regardless of the ages of the tokens).
For instance, if
$\place={\mbox{\tikz{\node[red-ball]{};}}}$
then $\marking_\place$ is the set of markings that have at least one token
in \tikz{\node[red-ball]{};}.

\paragraph{The Priced Coverability Problem.}
We will consider two variants of the cost problem,
the {\it Cost-Threshold} problem and the 
{\it Cost-Optimality} problem.
They are both characterized by an 
(i) \emph{initial}
marking $\initmarking$ that places a single token
(with age $0$) in a given initial place
$\initp$, and
(ii) a set of \emph{final} markings
$\marking_{\finalp}$ defined by a \emph{final} place $\finalp$.
In other words, we start from a marking where there is only one token
with age $0$ in $\initp$ and where all the other places are empty, 
and then consider
the cost of computations that takes us to
$\marking_{\finalp}$. 

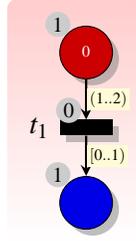
\begin{figure}
\center
\begin{tikzpicture}[
background rectangle/.style=
{top color=red!20,rounded corners}, show background rectangle,
>=stealth,every transition/.style={minimum width = 7mm,minimum height=2mm,inner sep=0mm,drop shadow,draw=none,fill=black}]

%%%%%%%%%%%%%%%%%%%  p1   %%%%%%%%%%%%%%%

  \node at (0,0) [red-place,label={[placeprice]135:{\scriptsize $1$}}] (p1){};

    \node at (p1) [] {\tiny
      \begin{tabular}{l}
        \white{$0$}
      \end{tabular}
    };

%%%%%%%%%%%%%%%%%%%  p2   %%%%%%%%%%%%%%%

  \node at (0mm,-20mm)[blue-place,label={[placeprice]135:{\scriptsize $1$}}] (p2){};

%%%%%%%%%%%%%%%%%%%  t1   %%%%%%%%%%%%%%%

     \node at (0mm, -10mm) [transition,label={180:$t_1$},label={[placeprice]135:{\scriptsize $0$}}] (t1){};

%%%%%%%%%%%%%%%%%%%  Edges  %%%%%%%%%%%%%%%%%%%%
   \draw [->,line width=0.3mm] (p1) -- node[right,intervalnode]{\tiny $(1..2)$} (t1);

   \draw [->,line width=0.3mm] (t1) -- node[right,intervalnode]{\tiny $[0..1)$}  (p2);

\end{tikzpicture}
\caption{A Simple {\sc Ptpn}.}
\label{simple:ptpn:figure}
\end{figure}

In the {\it Cost-Threshold} problem we ask the question whether
there is a computation starting from $\initmarking$ and reaching
a marking in $\marking_{\finalp}$ with a cost that is at most
$\threshold$
for a given threshold $\threshold \in \nat$.
In the {\it Cost-Optimality} problem, we want to compute
the \emph{optimal} (smallest) cost of reaching  $\marking_{\finalp}$
staring from $\initmarking$.
For given $\initmarking$ and
$\marking_{\finalp}$, the optimal cost of
reaching $\marking_{\finalp}$ from $\initmarking$ may not exist.
However, in \cite{abdulla2011computing},
we show that the infimum of the costs of all computations
is a natural number (or $\infty$ if $\marking_{\finalp}$ is not
reachable from $\initmarking$).
The situation is illustrated in Figure~\ref{simple:ptpn:figure}.
The optimal cost for putting a token in 
\tikz{\node[blue-ball]{};} can be made arbitrarily close to $1$ 
(but not equal to $1$).
In such a case, we simply define the optimal cost to be $1$.
In fact, the non-existence of an optimal cost has already
been observed for timed automata \cite{BCFL04}.

\section{Computations in $\delta$-Form}
\label{delta:section}
\noindent
In order to solve the 
Cost-Threshold and the 
Cost-Optimality  problems, it is sufficient
to consider computations of a certain form where the ages of 
all the tokens that appear in the computation are
arbitrarily close to
(within some small real number $\delta$ from) an integer.
Below, we assume a real number $\delta:0<\delta<0.2$.

\begin{figure}
\center
\begin{tikzpicture}[]

  \node[red-move-place,name=n1] {$7.93$};
  \node at ($(n1.east)+(1mm,0mm)$) [red-move-place,name=n2] {$1.06$};
  \node at ($(n2.east)+(1mm,0mm)$) [red-move-place,name=n3] {$2.02$};
  \node at ($(n3.east)+(1mm,0mm)$) [white-move-place,name=n4] {$2.00$};
  \node at ($(n4.east)+(1mm,0mm)$) [white-move-place,name=n5] {$0.97$};
  \node at ($(n5.east)+(1mm,0mm)$) [blue-move-place,name=n6] {$8.00$};
  \node at ($(n6.east)+(1mm,0mm)$) [green-move-place,name=n7] {$4.02$};
  \node at ($(n7.east)+(1mm,0mm)$) [green-move-place,name=n8] {$1.91$};
  \node at ($(n8.east)+(1mm,0mm)$) [orange-move-place,name=n9] {$2.03$};
  \node at ($(n9.east)+(1mm,0mm)$) [orange-move-place,name=n10] {$1.97$};
  \node at ($(n10.east)+(1mm,0mm)$) [orange-move-place,name=n11] {$4.02$};

\begin{pgfonlayer}{background}
\node[black-bg,fit=(n1) (n2) (n3) (n4) (n5) (n6) (n7) (n8) (n9) (n10) (n11)]{};
\end{pgfonlayer}

\end{tikzpicture}
\caption{A marking in $\delta$-form, $\delta=0.2$.}
\label{delta:marking:fig}
\end{figure}
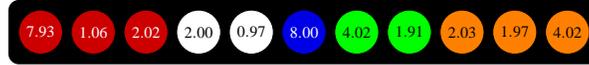

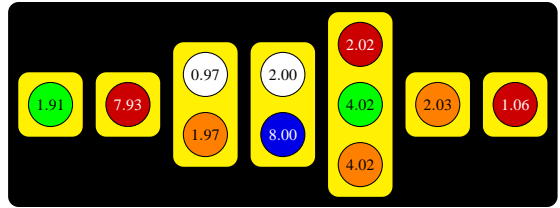
\begin{figure}
\center
\begin{tikzpicture}[]

  %\node[rounded corners,fill=black,rectangle,minimum width = 52mm,minimum height=21mm] at (21mm,0mm){};

  \begin{pgfonlayer}{foreground}
    \node at (0mm,0mm) [yellow-move-place,name=dummy] {$d.dd$};
    \node at (dummy.west) [green-move-place,name=p1] {$1.91$};
  \end{pgfonlayer}

  \node[yellow-bg,fit=(p1),name=c1]{};

 \begin{pgfonlayer}{foreground}
    \node at ($(dummy.east)+(4mm,0mm)$)  [yellow-move-place,name=dummy] {$d.dd$};
    \node at (dummy.west) [red-move-place,name=p1] {$7.93$};
  \end{pgfonlayer}

  \node[yellow-bg,fit=(p1),name=c2]{};

 \begin{pgfonlayer}{foreground}
    \node at ($(dummy.east)+(4mm,0mm)$)  [yellow-move-place,name=dummy] {$d.dd$};
    \node at ($(dummy.west)+(0mm,4mm)$)  [white-move-place,name=p1] {$0.97$};
    \node at ($(dummy.west)+(0mm,-4mm)$)  [orange-move-place,name=p2] {$1.97$};
  \end{pgfonlayer}

  \node[yellow-bg,fit=(p1) (p2),name=c3]{};

 \begin{pgfonlayer}{foreground}
    \node at ($(dummy.east)+(4mm,0mm)$)  [yellow-move-place,name=dummy] {$d.dd$};
    \node at ($(dummy.west)+(0mm,4mm)$) [white-move-place,name=p1] {$2.00$};
    \node at ($(dummy.west)+(0mm,-4mm)$)  [blue-move-place,name=p2] {$8.00$};
  \end{pgfonlayer}

  \node[yellow-bg,fit=(p1) (p2),name=c4]{};

 \begin{pgfonlayer}{foreground}
   \node at ($(dummy.east)+(4mm,0mm)$)  [yellow-move-place,name=dummy] {$d.dd$};
    \node at ($(dummy.west)+(0mm,8mm)$) [red-move-place,name=p1] {$2.02$};
    \node at (dummy.west)  [green-move-place,name=p2] {$4.02$};
    \node at ($(dummy.west)+(0mm,-8mm)$) [orange-move-place,name=p3] {$4.02$};
  \end{pgfonlayer}

  \node[yellow-bg,fit=(p1) (p2) (p3),name=c5]{};

 \begin{pgfonlayer}{foreground}
   \node at ($(dummy.east)+(4mm,0mm)$)  [yellow-move-place,name=dummy] {$d.dd$};
     \node at (dummy.west)  [orange-move-place,name=p1] {$2.03$};
   \end{pgfonlayer}

  \node[yellow-bg,fit=(p1),name=c6]{};

\begin{pgfonlayer}{foreground}
   \node at ($(dummy.east)+(4mm,0mm)$)  [yellow-move-place,name=dummy] {$d.dd$};
     \node at (dummy.west)  [red-move-place,name=p1] {$1.06$};
   \end{pgfonlayer}

  \node[yellow-bg,fit=(p1),name=c7]{};

  \begin{pgfonlayer}{background}
    \node[black-bg,fit=(c1) (c2) (c3) (c4) (c5) (c6) (c7)]{};
  \end{pgfonlayer}

 \end{tikzpicture}
\caption{The partitioning the marking in Figure~\ref{delta:marking:fig}.}
\label{delta:marking:partition:fig}
\end{figure}

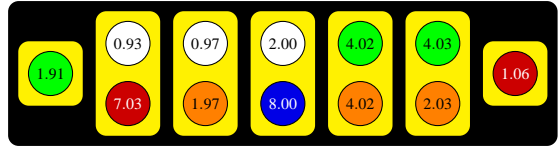
\begin{figure}
\center
\begin{tikzpicture}[]

  %\node[rounded corners,fill=black,rectangle,minimum width = 52mm,minimum height=21mm] at (21mm,0mm){};

  \begin{pgfonlayer}{foreground}
    \node at (0mm,0mm) [yellow-move-place,name=dummy] {$d.dd$};
    \node at (dummy.west) [green-move-place,name=p1] {$1.91$};
  \end{pgfonlayer}

  \node[yellow-bg,fit=(p1),name=c1]{};

 \begin{pgfonlayer}{foreground}
    \node at ($(dummy.east)+(4mm,0mm)$)  [yellow-move-place,name=dummy] {$d.dd$};
    \node at ($(dummy.west)+(0mm,4mm)$)  [white-move-place,name=p1] {$0.93$};
    \node at ($(dummy.west)+(0mm,-4mm)$)  [red-move-place,name=p2] {$7.03$};
  \end{pgfonlayer}

  \node[yellow-bg,fit=(p1) (p2),name=c2]{};

 \begin{pgfonlayer}{foreground}
    \node at ($(dummy.east)+(4mm,0mm)$)  [yellow-move-place,name=dummy] {$d.dd$};
    \node at ($(dummy.west)+(0mm,4mm)$)  [white-move-place,name=p1] {$0.97$};
    \node at ($(dummy.west)+(0mm,-4mm)$)  [orange-move-place,name=p2] {$1.97$};
  \end{pgfonlayer}

  \node[yellow-bg,fit=(p1) (p2),name=c3]{};

 \begin{pgfonlayer}{foreground}
    \node at ($(dummy.east)+(4mm,0mm)$)  [yellow-move-place,name=dummy] {$d.dd$};
    \node at ($(dummy.west)+(0mm,4mm)$) [white-move-place,name=p1] {$2.00$};
    \node at ($(dummy.west)+(0mm,-4mm)$)  [blue-move-place,name=p2] {$8.00$};
  \end{pgfonlayer}

  \node[yellow-bg,fit=(p1) (p2),name=c4]{};

 \begin{pgfonlayer}{foreground}
   \node at ($(dummy.east)+(4mm,0mm)$)  [yellow-move-place,name=dummy] {$d.dd$};
    \node at ($(dummy.west)+(0mm,4mm)$) [green-move-place,name=p1] {$4.02$};
    \node at ($(dummy.west)+(0mm,-4mm)$) [orange-move-place,name=p2] {$4.02$};
  \end{pgfonlayer}

  \node[yellow-bg,fit=(p1) (p2),name=c5]{};

 \begin{pgfonlayer}{foreground}
   \node at ($(dummy.east)+(4mm,0mm)$)  [yellow-move-place,name=dummy] {$d.dd$};
    \node at ($(dummy.west)+(0mm,4mm)$) [green-move-place,name=p1] {$4.03$};
    \node at ($(dummy.west)+(0mm,-4mm)$) [orange-move-place,name=p2] {$2.03$};
   \end{pgfonlayer}

  \node[yellow-bg,fit=(p1) (p2),name=c6]{};

\begin{pgfonlayer}{foreground}
   \node at ($(dummy.east)+(4mm,0mm)$)  [yellow-move-place,name=dummy] {$d.dd$};
     \node at (dummy.west)  [red-move-place,name=p1] {$1.06$};
   \end{pgfonlayer}

  \node[yellow-bg,fit=(p1),name=c7]{};

  \begin{pgfonlayer}{background}
    \node[black-bg,fit=(c1) (c2) (c3) (c4) (c5) (c6) (c7)]{};
  \end{pgfonlayer}

\end{tikzpicture}
\caption{An application in $\delta$-form ($\delta=0.2$)
 of $t_1$ on the marking of 
Figure~\ref{delta:marking:fig}. The two new tokens have fractional parts
that are equal to $0.93$ resp.\ $0.03$.}
\label{disc:detailed:fig}
\end{figure}

\paragraph{$\delta$-Markings.}
A marking $\marking$ is said to be in $\delta$-form
(Figure~\ref{delta:marking:fig})
if any fractional part of the age of a token appearing in 
$\marking$ is either smaller than $\delta$ or larger than $1-\delta$.
We decompose a $\delta$-marking
into submarkings such that in every submarking the 
fractional parts (but not necessarily the integer parts) 
of the token ages are
identical. We then arrange these submarkings in a sequence  
$
\marking_{-m}, \dots , \marking_{-1}, \marking_0,
\marking_1,\dots,\marking_n
$
such that $\marking_{-m}, \dots , \marking_{-1}$ contain tokens with fractional 
parts $\ge \delta$ in increasing order,
$\marking_0$ contains the tokens with fractional part zero,
and $\marking_1,\dots,\marking_n$ contain 
tokens with fractional 
parts $< \delta$ in increasing order.
Figure~\ref{delta:marking:partition:fig} shows that partitioning of
the marking $\marking$ in Figure~\ref{delta:marking:fig}.
More precisely, We start with the token with the high fractional parts, namely
$0.91$ (one token in \tikz{\node[green-ball]{};}), followed by
$0.93$ (one token in \tikz{\node[red-ball]{};}), followed by
$0.97$ (one token in \tikz{\node[white-ball]{};} and one token in \tikz{\node[orange-ball]{};}).
Furthermore, there are two tokens with zero fractional parts
(one token in \tikz{\node[white-ball]{};} and one token in \tikz{\node[blue-ball]{};}).
Finally, we consider the tokens with low fractional parts, 
namely
$0.02$ (one token in \tikz{\node[red-ball]{};},
one token in \tikz{\node[green-ball]{};}, and one token in \tikz{\node[orange-ball]{};}),
followed by
$0.03$ (one token in \tikz{\node[orange-ball]{};}), followed by
$00.7$ (one token in \tikz{\node[red-ball]{};}).
\paragraph{Computations in $\delta$-form.}
The occurrence of a discrete transition $t$ is said to be in $\delta$-form if 
the ages of the newly generated tokens are  
close to an integer (i.e., within distance $\delta$). This is not a property
of the transition $t$ as such, but a property of its occurrence.
Figure~\ref{disc:detailed:fig} shows the result of an occurrence of
$t_1$ in $\delta$-form (with $\delta=0.2$) on the marking of 
Figure~\ref{delta:marking:partition:fig}.

A computation is in $\delta$-form
if:
\begin{enumerate}
\item 
Every occurrence of a discrete transition 
is in $\delta$-form, and 
\item
For every timed transition, the delay is
either in the interval $(0:\delta)$ or in the interval $x \in (1-\delta:1)$.
\end{enumerate}

\paragraph{Detailed Timed Transitions.}
We say that a timed transition (from a marking $\marking$)
is {\em detailed} iff at most one fractional part of any token in $\marking$
changes its status about reaching or exceeding the next integer value.
Figures~\ref{detailed:fig}~and~\ref{detailed:cont:fig}
 show some steps in a detailed
computation.
In the first transition, time passes by a positive amount but not sufficiently
long to make any tokens with positive fractional parts to 
increase to the next integer.
More precisely, the time delay is $0.01$ which means that two tokens
in 
\tikz{\node[white-ball]{};} 
and
\tikz{\node[blue-ball]{};}
that have zero fractional parts, will now have positive fractional parts 
($0.1$).
On the other hand, the two tokens in
\tikz{\node[white-ball]{};}
and
\tikz{\node[orange-ball]{};}
that have the highest fractional parts ($0.8$)
will not cross to the next integer
(their ages will now be $0.98$ and $1.98$
respectively).
\begin{figure}
\center
\begin{tikzpicture}[]

\begin{scope}[xshift=0mm,yshift=0mm]

  \begin{pgfonlayer}{foreground}
    \node at (0mm,0mm) [yellow-move-place,name=dummy] {$d.dd$};
    \node at (dummy.west) [green-move-place,name=p1] {$1.91$};
  \end{pgfonlayer}

  \node[yellow-bg,fit=(p1),name=c1]{};

 \begin{pgfonlayer}{foreground}
    \node at ($(dummy.east)+(4mm,0mm)$)  [yellow-move-place,name=dummy] {$d.dd$};
    \node at (dummy.west) [red-move-place,name=p1] {$7.93$};
  \end{pgfonlayer}

  \node[yellow-bg,fit=(p1),name=c2]{};

 \begin{pgfonlayer}{foreground}
    \node at ($(dummy.east)+(4mm,0mm)$)  [yellow-move-place,name=dummy] {$d.dd$};
    \node at ($(dummy.west)+(0mm,4mm)$)  [white-move-place,name=p1] {$0.97$};
    \node at ($(dummy.west)+(0mm,-4mm)$)  [orange-move-place,name=p2] {$1.97$};
  \end{pgfonlayer}

  \node[yellow-bg,fit=(p1) (p2),name=c3]{};

 \begin{pgfonlayer}{foreground}
    \node at ($(dummy.east)+(4mm,0mm)$)  [yellow-move-place,name=dummy] {$d.dd$};
    \node at ($(dummy.west)+(0mm,4mm)$) [white-move-place,name=p1] {$2.00$};
    \node at ($(dummy.west)+(0mm,-4mm)$)  [blue-move-place,name=p2] {$8.00$};
  \end{pgfonlayer}

  \node[yellow-bg,fit=(p1) (p2),name=c4]{};

 \begin{pgfonlayer}{foreground}
   \node at ($(dummy.east)+(4mm,0mm)$)  [yellow-move-place,name=dummy] {$d.dd$};
    \node at ($(dummy.west)+(0mm,8mm)$) [red-move-place,name=p1] {$2.02$};
    \node at (dummy.west)  [green-move-place,name=p2] {$4.02$};
    \node at ($(dummy.west)+(0mm,-8mm)$) [orange-move-place,name=p3] {$4.02$};
  \end{pgfonlayer}

  \node[yellow-bg,fit=(p1) (p2) (p3),name=c5]{};

 \begin{pgfonlayer}{foreground}
   \node at ($(dummy.east)+(4mm,0mm)$)  [yellow-move-place,name=dummy] {$d.dd$};
     \node at (dummy.west)  [orange-move-place,name=p1] {$2.03$};
   \end{pgfonlayer}

  \node[yellow-bg,fit=(p1),name=c6]{};

\begin{pgfonlayer}{foreground}
   \node at ($(dummy.east)+(4mm,0mm)$)  [yellow-move-place,name=dummy] {$d.dd$};
     \node at (dummy.west)  [red-move-place,name=p1] {$1.06$};
   \end{pgfonlayer}

  \node[yellow-bg,fit=(p1),name=c7]{};

  \begin{pgfonlayer}{background}
    \node[black-bg,fit=(c1) (c2) (c3) (c4) (c5) (c6) (c7)]{};
  \end{pgfonlayer}

  \draw[->,line width=1pt] ($(c7.east)+(4mm,0mm)$) -- ($(c7.east)+(17mm,0mm)$);

\end{scope}

\begin{scope}[xshift= 0mm,yshift=-35mm]

  \begin{pgfonlayer}{foreground}
    \node at (0mm,0mm) [yellow-move-place,name=dummy] {$d.dd$};
    \node at (dummy.west) [green-move-place,name=p1] {$1.92$};
  \end{pgfonlayer}

  \node[yellow-bg,fit=(p1),name=c1]{};

 \begin{pgfonlayer}{foreground}
    \node at ($(dummy.east)+(4mm,0mm)$)  [yellow-move-place,name=dummy] {$d.dd$};
    \node at (dummy.west) [red-move-place,name=p1] {$7.94$};
  \end{pgfonlayer}

  \node[yellow-bg,fit=(p1),name=c2]{};

 \begin{pgfonlayer}{foreground}
    \node at ($(dummy.east)+(4mm,0mm)$)  [yellow-move-place,name=dummy] {$d.dd$};
    \node at ($(dummy.west)+(0mm,4mm)$)  [white-move-place,name=p1] {$0.98$};
    \node at ($(dummy.west)+(0mm,-4mm)$)  [orange-move-place,name=p2] {$1.98$};
  \end{pgfonlayer}

  \node[yellow-bg,fit=(p1) (p2),name=c3]{};

 \begin{pgfonlayer}{foreground}
    \node at ($(dummy.east)+(4mm,0mm)$)  [yellow-move-place,name=dummy] {$d.dd$};
    \end{pgfonlayer}

  \node[yellow-bg,fit=(dummy),name=c3.5]{};

 \begin{pgfonlayer}{foreground}
    \node at ($(dummy.east)+(4mm,0mm)$)  [yellow-move-place,name=dummy] {$d.dd$};
    \node at ($(dummy.west)+(0mm,4mm)$) [white-move-place,name=p1] {$2.01$};
    \node at ($(dummy.west)+(0mm,-4mm)$)  [blue-move-place,name=p2] {$8.01$};
  \end{pgfonlayer}

  \node[yellow-bg,fit=(p1) (p2),name=c4]{};

 \begin{pgfonlayer}{foreground}
   \node at ($(dummy.east)+(4mm,0mm)$)  [yellow-move-place,name=dummy] {$d.dd$};
    \node at ($(dummy.west)+(0mm,8mm)$) [red-move-place,name=p1] {$2.03$};
    \node at (dummy.west)  [green-move-place,name=p2] {$4.03$};
    \node at ($(dummy.west)+(0mm,-8mm)$) [orange-move-place,name=p3] {$4.03$};
  \end{pgfonlayer}

  \node[yellow-bg,fit=(p1) (p2) (p3),name=c5]{};

 \begin{pgfonlayer}{foreground}
   \node at ($(dummy.east)+(4mm,0mm)$)  [yellow-move-place,name=dummy] {$d.dd$};
     \node at (dummy.west)  [orange-move-place,name=p1] {$2.04$};
   \end{pgfonlayer}

  \node[yellow-bg,fit=(p1),name=c6]{};

\begin{pgfonlayer}{foreground}
   \node at ($(dummy.east)+(4mm,0mm)$)  [yellow-move-place,name=dummy] {$d.dd$};
     \node at (dummy.west)  [red-move-place,name=p1] {$1.07$};
   \end{pgfonlayer}

  \node[yellow-bg,fit=(p1),name=c7]{};

  \begin{pgfonlayer}{background}
    \node[black-bg,fit=(c1) (c2) (c3) (c4) (c5) (c6) (c7)]{};
  \end{pgfonlayer}

  \draw[->,line width=1pt] ($(c7.east)+(4mm,0mm)$) -- ($(c7.east)+(17mm,0mm)$);

\end{scope}

\begin{scope}[xshift=0mm,yshift=-70mm]

  \begin{pgfonlayer}{foreground}
    \node at (0mm,0mm) [yellow-move-place,name=dummy] {$d.dd$};
    \node at (dummy.west) [green-move-place,name=p1] {$1.94$};
  \end{pgfonlayer}

  \node[yellow-bg,fit=(p1),name=c1]{};

 \begin{pgfonlayer}{foreground}
    \node at ($(dummy.east)+(4mm,0mm)$)  [yellow-move-place,name=dummy] {$d.dd$};
    \node at (dummy.west) [red-move-place,name=p1] {$7.96$};
  \end{pgfonlayer}

  \node[yellow-bg,fit=(p1),name=c2]{};

 \begin{pgfonlayer}{foreground}
    \node at ($(dummy.east)+(4mm,0mm)$)  [yellow-move-place,name=dummy] {$d.dd$};
    \node at ($(dummy.west)+(0mm,4mm)$)  [white-move-place,name=p1] {$1.00$};
    \node at ($(dummy.west)+(0mm,-4mm)$)  [orange-move-place,name=p2] {$2.00$};
  \end{pgfonlayer}

  \node[yellow-bg,fit=(p1) (p2),name=c3]{};

 \begin{pgfonlayer}{foreground}
    \node at ($(dummy.east)+(4mm,0mm)$)  [yellow-move-place,name=dummy] {$d.dd$};
    \node at ($(dummy.west)+(0mm,4mm)$) [white-move-place,name=p1] {$2.03$};
    \node at ($(dummy.west)+(0mm,-4mm)$)  [blue-move-place,name=p2] {$8.03$};
  \end{pgfonlayer}

  \node[yellow-bg,fit=(p1) (p2),name=c4]{};

 \begin{pgfonlayer}{foreground}
   \node at ($(dummy.east)+(4mm,0mm)$)  [yellow-move-place,name=dummy] {$d.dd$};
    \node at ($(dummy.west)+(0mm,8mm)$) [red-move-place,name=p1] {$2.05$};
    \node at (dummy.west)  [green-move-place,name=p2] {$4.05$};
    \node at ($(dummy.west)+(0mm,-8mm)$) [orange-move-place,name=p3] {$4.05$};
  \end{pgfonlayer}

  \node[yellow-bg,fit=(p1) (p2) (p3),name=c5]{};

 \begin{pgfonlayer}{foreground}
   \node at ($(dummy.east)+(4mm,0mm)$)  [yellow-move-place,name=dummy] {$d.dd$};
     \node at (dummy.west)  [orange-move-place,name=p1] {$2.06$};
   \end{pgfonlayer}

  \node[yellow-bg,fit=(p1),name=c6]{};

\begin{pgfonlayer}{foreground}
   \node at ($(dummy.east)+(4mm,0mm)$)  [yellow-move-place,name=dummy] {$d.dd$};
     \node at (dummy.west)  [red-move-place,name=p1] {$1.09$};
   \end{pgfonlayer}

  \node[yellow-bg,fit=(p1),name=c7]{};

  \begin{pgfonlayer}{background}
    \node[black-bg,fit=(c1) (c2) (c3) (c4) (c5) (c6) (c7)]{};
  \end{pgfonlayer}

  \draw[->,line width=1pt] ($(c7.east)+(4mm,0mm)$) -- ($(c7.east)+(17mm,0mm)$);

 \end{scope}

\begin{scope}[xshift= 0mm,yshift=-105mm]

  \begin{pgfonlayer}{foreground}
    \node at (0mm,0mm) [yellow-move-place,name=dummy] {$d.dd$};
    \node at (dummy.west) [green-move-place,name=p1] {$1.97$};
  \end{pgfonlayer}

  \node[yellow-bg,fit=(p1),name=c1]{};

 \begin{pgfonlayer}{foreground}
    \node at ($(dummy.east)+(4mm,0mm)$)  [yellow-move-place,name=dummy] {$d.dd$};
    \node at (dummy.west) [red-move-place,name=p1] {$7.99$};
  \end{pgfonlayer}

  \node[yellow-bg,fit=(p1),name=c2]{};

 \begin{pgfonlayer}{foreground}
    \node at ($(dummy.east)+(4mm,0mm)$)  [yellow-move-place,name=dummy] {$d.dd$};
    \end{pgfonlayer}

  \node[yellow-bg,fit=(dummy),name=c2.5]{};

 \begin{pgfonlayer}{foreground}
    \node at ($(dummy.east)+(4mm,0mm)$)  [yellow-move-place,name=dummy] {$d.dd$};
    \node at ($(dummy.west)+(0mm,4mm)$)  [white-move-place,name=p1] {$1.03$};
    \node at ($(dummy.west)+(0mm,-4mm)$)  [orange-move-place,name=p2] {$2.03$};
  \end{pgfonlayer}

  \node[yellow-bg,fit=(p1) (p2),name=c3]{};

 \begin{pgfonlayer}{foreground}
    \node at ($(dummy.east)+(4mm,0mm)$)  [yellow-move-place,name=dummy] {$d.dd$};
    \node at ($(dummy.west)+(0mm,4mm)$) [white-move-place,name=p1] {$2.06$};
    \node at ($(dummy.west)+(0mm,-4mm)$)  [blue-move-place,name=p2] {$8.06$};
  \end{pgfonlayer}

  \node[yellow-bg,fit=(p1) (p2),name=c4]{};

 \begin{pgfonlayer}{foreground}
   \node at ($(dummy.east)+(4mm,0mm)$)  [yellow-move-place,name=dummy] {$d.dd$};
    \node at ($(dummy.west)+(0mm,8mm)$) [red-move-place,name=p1] {$2.08$};
    \node at (dummy.west)  [green-move-place,name=p2] {$4.08$};
    \node at ($(dummy.west)+(0mm,-8mm)$) [orange-move-place,name=p3] {$4.08$};
  \end{pgfonlayer}

  \node[yellow-bg,fit=(p1) (p2) (p3),name=c5]{};

 \begin{pgfonlayer}{foreground}
   \node at ($(dummy.east)+(4mm,0mm)$)  [yellow-move-place,name=dummy] {$d.dd$};
     \node at (dummy.west)  [orange-move-place,name=p1] {$2.09$};
   \end{pgfonlayer}

  \node[yellow-bg,fit=(p1),name=c6]{};

\begin{pgfonlayer}{foreground}
   \node at ($(dummy.east)+(4mm,0mm)$)  [yellow-move-place,name=dummy] {$d.dd$};
     \node at (dummy.west)  [red-move-place,name=p1] {$1.12$};
   \end{pgfonlayer}

  \node[yellow-bg,fit=(p1),name=c7]{};

  \begin{pgfonlayer}{background}
    \node[black-bg,fit=(c1) (c2) (c3) (c4) (c5) (c6) (c7)]{};
  \end{pgfonlayer}

  \draw[->,line width=1pt] ($(c7.east)+(4mm,0mm)$) -- ($(c7.east)+(17mm,0mm)$);

\end{scope}

\begin{scope}[xshift=0mm,yshift=-140mm]

  \begin{pgfonlayer}{foreground}
    \node at (0mm,0mm) [yellow-move-place,name=dummy] {$d.dd$};
    \node at (dummy.west) [green-move-place,name=p1] {$1.98$};
  \end{pgfonlayer}

  \node[yellow-bg,fit=(p1),name=c1]{};

 \begin{pgfonlayer}{foreground}
    \node at ($(dummy.east)+(4mm,0mm)$)  [yellow-move-place,name=dummy] {$d.dd$};
    \node at (dummy.west) [red-move-place,name=p1] {$8.00$};
  \end{pgfonlayer}

  \node[yellow-bg,fit=(p1),name=c2]{};

 \begin{pgfonlayer}{foreground}
    \node at ($(dummy.east)+(4mm,0mm)$)  [yellow-move-place,name=dummy] {$d.dd$};
    \node at ($(dummy.west)+(0mm,4mm)$)  [white-move-place,name=p1] {$1.04$};
    \node at ($(dummy.west)+(0mm,-4mm)$)  [orange-move-place,name=p2] {$2.04$};
  \end{pgfonlayer}

  \node[yellow-bg,fit=(p1) (p2),name=c3]{};

 \begin{pgfonlayer}{foreground}
    \node at ($(dummy.east)+(4mm,0mm)$)  [yellow-move-place,name=dummy] {$d.dd$};
    \node at ($(dummy.west)+(0mm,4mm)$) [white-move-place,name=p1] {$2.07$};
    \node at ($(dummy.west)+(0mm,-4mm)$)  [blue-move-place,name=p2] {$8.07$};
  \end{pgfonlayer}

  \node[yellow-bg,fit=(p1) (p2),name=c4]{};

 \begin{pgfonlayer}{foreground}
   \node at ($(dummy.east)+(4mm,0mm)$)  [yellow-move-place,name=dummy] {$d.dd$};
    \node at ($(dummy.west)+(0mm,8mm)$) [red-move-place,name=p1] {$2.09$};
    \node at (dummy.west)  [green-move-place,name=p2] {$4.09$};
    \node at ($(dummy.west)+(0mm,-8mm)$) [orange-move-place,name=p3] {$4.09$};
  \end{pgfonlayer}

  \node[yellow-bg,fit=(p1) (p2) (p3),name=c5]{};

 \begin{pgfonlayer}{foreground}
   \node at ($(dummy.east)+(4mm,0mm)$)  [yellow-move-place,name=dummy] {$d.dd$};
     \node at (dummy.west)  [orange-move-place,name=p1] {$2.10$};
   \end{pgfonlayer}

  \node[yellow-bg,fit=(p1),name=c6]{};

\begin{pgfonlayer}{foreground}
   \node at ($(dummy.east)+(4mm,0mm)$)  [yellow-move-place,name=dummy] {$d.dd$};
     \node at (dummy.west)  [red-move-place,name=p1] {$1.13$};
   \end{pgfonlayer}

  \node[yellow-bg,fit=(p1),name=c7]{};

  \begin{pgfonlayer}{background}
    \node[black-bg,fit=(c1) (c2) (c3) (c4) (c5) (c6) (c7)]{};
  \end{pgfonlayer}

  \draw[->,line width=1pt] ($(c7.east)+(4mm,0mm)$) -- ($(c7.east)+(17mm,0mm)$);

 \end{scope}

\end{tikzpicture}
\caption{Detailed timed transitions for $\delta=0.2$.}
\label{detailed:fig}
\end{figure}
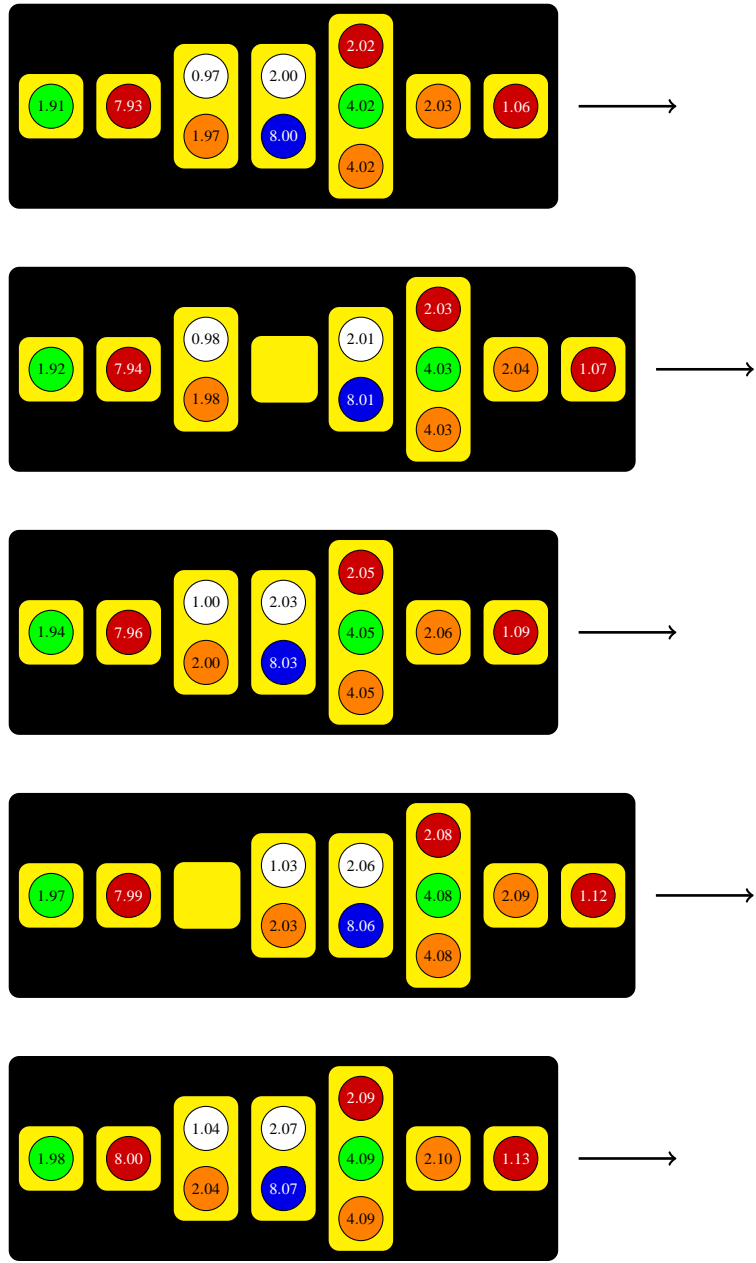
\begin{figure}
\center
\begin{tikzpicture}[]

\begin{scope}[xshift=0mm,yshift=0mm]

  \begin{pgfonlayer}{foreground}
    \node at (0mm,0mm) [yellow-move-place,name=dummy] {$d.dd$};
    \node at (dummy.west) [green-move-place,name=p1] {$1.99$};
  \end{pgfonlayer}

  \node[yellow-bg,fit=(p1),name=c1]{};

 \begin{pgfonlayer}{foreground}
    \node at ($(dummy.east)+(4mm,0mm)$)  [yellow-move-place,name=dummy] {$d.dd$};
    \end{pgfonlayer}

  \node[yellow-bg,fit=(dummy),name=c1.5]{};

 \begin{pgfonlayer}{foreground}
    \node at ($(dummy.east)+(4mm,0mm)$)  [yellow-move-place,name=dummy] {$d.dd$};
    \node at (dummy.west) [red-move-place,name=p1] {$8.01$};
  \end{pgfonlayer}

  \node[yellow-bg,fit=(p1),name=c2]{};

 \begin{pgfonlayer}{foreground}
    \node at ($(dummy.east)+(4mm,0mm)$)  [yellow-move-place,name=dummy] {$d.dd$};
    \node at ($(dummy.west)+(0mm,4mm)$)  [white-move-place,name=p1] {$1.05$};
    \node at ($(dummy.west)+(0mm,-4mm)$)  [orange-move-place,name=p2] {$2.05$};
  \end{pgfonlayer}

  \node[yellow-bg,fit=(p1) (p2),name=c3]{};

 \begin{pgfonlayer}{foreground}
    \node at ($(dummy.east)+(4mm,0mm)$)  [yellow-move-place,name=dummy] {$d.dd$};
    \node at ($(dummy.west)+(0mm,4mm)$) [white-move-place,name=p1] {$2.08$};
    \node at ($(dummy.west)+(0mm,-4mm)$)  [blue-move-place,name=p2] {$8.08$};
  \end{pgfonlayer}

  \node[yellow-bg,fit=(p1) (p2),name=c4]{};

 \begin{pgfonlayer}{foreground}
   \node at ($(dummy.east)+(4mm,0mm)$)  [yellow-move-place,name=dummy] {$d.dd$};
    \node at ($(dummy.west)+(0mm,8mm)$) [red-move-place,name=p1] {$2.10$};
    \node at (dummy.west)  [green-move-place,name=p2] {$4.10$};
    \node at ($(dummy.west)+(0mm,-8mm)$) [orange-move-place,name=p3] {$4.10$};
  \end{pgfonlayer}

  \node[yellow-bg,fit=(p1) (p2) (p3),name=c5]{};

 \begin{pgfonlayer}{foreground}
   \node at ($(dummy.east)+(4mm,0mm)$)  [yellow-move-place,name=dummy] {$d.dd$};
     \node at (dummy.west)  [orange-move-place,name=p1] {$2.11$};
   \end{pgfonlayer}

  \node[yellow-bg,fit=(p1),name=c6]{};

\begin{pgfonlayer}{foreground}
   \node at ($(dummy.east)+(4mm,0mm)$)  [yellow-move-place,name=dummy] {$d.dd$};
     \node at (dummy.west)  [red-move-place,name=p1] {$1.14$};
   \end{pgfonlayer}

  \node[yellow-bg,fit=(p1),name=c7]{};

  \begin{pgfonlayer}{background}
    \node[black-bg,fit=(c1) (c2) (c3) (c4) (c5) (c6) (c7)]{};
  \end{pgfonlayer}

  \draw[->,line width=1pt] ($(c7.east)+(4mm,0mm)$) -- ($(c7.east)+(17mm,0mm)$);

\end{scope}

\begin{scope}[xshift=0mm,yshift=-35mm]

  \begin{pgfonlayer}{foreground}
    \node at (0mm,0mm) [yellow-move-place,name=dummy] {$d.dd$};
    \node at (dummy.west) [green-move-place,name=p1] {$2.00$};
  \end{pgfonlayer}

  \node[yellow-bg,fit=(p1),name=c1]{};

 \begin{pgfonlayer}{foreground}
    \node at ($(dummy.east)+(4mm,0mm)$)  [yellow-move-place,name=dummy] {$d.dd$};
    \node at (dummy.west) [red-move-place,name=p1] {$8.02$};
  \end{pgfonlayer}

  \node[yellow-bg,fit=(p1),name=c2]{};

 \begin{pgfonlayer}{foreground}
    \node at ($(dummy.east)+(4mm,0mm)$)  [yellow-move-place,name=dummy] {$d.dd$};
    \node at ($(dummy.west)+(0mm,4mm)$)  [white-move-place,name=p1] {$1.06$};
    \node at ($(dummy.west)+(0mm,-4mm)$)  [orange-move-place,name=p2] {$2.06$};
  \end{pgfonlayer}

  \node[yellow-bg,fit=(p1) (p2),name=c3]{};

 \begin{pgfonlayer}{foreground}
    \node at ($(dummy.east)+(4mm,0mm)$)  [yellow-move-place,name=dummy] {$d.dd$};
    \node at ($(dummy.west)+(0mm,4mm)$) [white-move-place,name=p1] {$2.09$};
    \node at ($(dummy.west)+(0mm,-4mm)$)  [blue-move-place,name=p2] {$8.09$};
  \end{pgfonlayer}

  \node[yellow-bg,fit=(p1) (p2),name=c4]{};

 \begin{pgfonlayer}{foreground}
   \node at ($(dummy.east)+(4mm,0mm)$)  [yellow-move-place,name=dummy] {$d.dd$};
    \node at ($(dummy.west)+(0mm,8mm)$) [red-move-place,name=p1] {$2.11$};
    \node at (dummy.west)  [green-move-place,name=p2] {$4.11$};
    \node at ($(dummy.west)+(0mm,-8mm)$) [orange-move-place,name=p3] {$4.11$};
  \end{pgfonlayer}

  \node[yellow-bg,fit=(p1) (p2) (p3),name=c5]{};

 \begin{pgfonlayer}{foreground}
   \node at ($(dummy.east)+(4mm,0mm)$)  [yellow-move-place,name=dummy] {$d.dd$};
     \node at (dummy.west)  [orange-move-place,name=p1] {$2.12$};
   \end{pgfonlayer}

  \node[yellow-bg,fit=(p1),name=c6]{};

\begin{pgfonlayer}{foreground}
   \node at ($(dummy.east)+(4mm,0mm)$)  [yellow-move-place,name=dummy] {$d.dd$};
     \node at (dummy.west)  [red-move-place,name=p1] {$1.15$};
   \end{pgfonlayer}

  \node[yellow-bg,fit=(p1),name=c7]{};

  \begin{pgfonlayer}{background}
    \node[black-bg,fit=(c1) (c2) (c3) (c4) (c5) (c6) (c7)]{};
  \end{pgfonlayer}

  \draw[->,line width=1pt] ($(c7.east)+(4mm,0mm)$) -- ($(c7.east)+(17mm,0mm)$);

 \end{scope}

\begin{scope}[xshift=0mm,yshift=-70mm]

 \begin{pgfonlayer}{foreground}
    \node at (0,0)  [yellow-move-place,name=dummy] {$d.dd$};
    \end{pgfonlayer}

  \node[yellow-bg,fit=(dummy),name=c05]{};

  \begin{pgfonlayer}{foreground}
    \node at ($(dummy.east)+(4mm,0mm)$) [yellow-move-place,name=dummy] {$d.dd$};
    \node at (dummy.west) [green-move-place,name=p1] {$2.03$};
  \end{pgfonlayer}

  \node[yellow-bg,fit=(p1),name=c1]{};

 \begin{pgfonlayer}{foreground}
    \node at ($(dummy.east)+(4mm,0mm)$)  [yellow-move-place,name=dummy] {$d.dd$};
    \node at (dummy.west) [red-move-place,name=p1] {$8.05$};
  \end{pgfonlayer}

  \node[yellow-bg,fit=(p1),name=c2]{};

 \begin{pgfonlayer}{foreground}
    \node at ($(dummy.east)+(4mm,0mm)$)  [yellow-move-place,name=dummy] {$d.dd$};
    \node at ($(dummy.west)+(0mm,4mm)$)  [white-move-place,name=p1] {$1.09$};
    \node at ($(dummy.west)+(0mm,-4mm)$)  [orange-move-place,name=p2] {$2.09$};
  \end{pgfonlayer}

  \node[yellow-bg,fit=(p1) (p2),name=c3]{};

 \begin{pgfonlayer}{foreground}
    \node at ($(dummy.east)+(4mm,0mm)$)  [yellow-move-place,name=dummy] {$d.dd$};
    \node at ($(dummy.west)+(0mm,4mm)$) [white-move-place,name=p1] {$2.12$};
    \node at ($(dummy.west)+(0mm,-4mm)$)  [blue-move-place,name=p2] {$8.12$};
  \end{pgfonlayer}

  \node[yellow-bg,fit=(p1) (p2),name=c4]{};

 \begin{pgfonlayer}{foreground}
   \node at ($(dummy.east)+(4mm,0mm)$)  [yellow-move-place,name=dummy] {$d.dd$};
    \node at ($(dummy.west)+(0mm,8mm)$) [red-move-place,name=p1] {$2.14$};
    \node at (dummy.west)  [green-move-place,name=p2] {$4.14$};
    \node at ($(dummy.west)+(0mm,-8mm)$) [orange-move-place,name=p3] {$4.14$};
  \end{pgfonlayer}

  \node[yellow-bg,fit=(p1) (p2) (p3),name=c5]{};

 \begin{pgfonlayer}{foreground}
   \node at ($(dummy.east)+(4mm,0mm)$)  [yellow-move-place,name=dummy] {$d.dd$};
     \node at (dummy.west)  [orange-move-place,name=p1] {$2.15$};
   \end{pgfonlayer}

  \node[yellow-bg,fit=(p1),name=c6]{};

\begin{pgfonlayer}{foreground}
   \node at ($(dummy.east)+(4mm,0mm)$)  [yellow-move-place,name=dummy] {$d.dd$};
     \node at (dummy.west)  [red-move-place,name=p1] {$1.18$};
   \end{pgfonlayer}

  \node[yellow-bg,fit=(p1),name=c7]{};

  \begin{pgfonlayer}{background}
    \node[black-bg,fit= (c05) (c1) (c2) (c3) (c4) (c5) (c6) (c7)]{};
  \end{pgfonlayer}

  \draw[->,line width=1pt] ($(c7.east)+(4mm,0mm)$) -- ($(c7.east)+(17mm,0mm)$);

\end{scope}

\begin{scope}[xshift=0mm,yshift=-105mm]

  \begin{pgfonlayer}{foreground}
    \node at (0,0) [yellow-move-place,name=dummy] {$d.dd$};
    \node at (dummy.west) [green-move-place,name=p1] {$2.81$};
  \end{pgfonlayer}

  \node[yellow-bg,fit=(p1),name=c1]{};

 \begin{pgfonlayer}{foreground}
    \node at ($(dummy.east)+(4mm,0mm)$)  [yellow-move-place,name=dummy] {$d.dd$};
    \node at (dummy.west) [red-move-place,name=p1] {$8.83$};
  \end{pgfonlayer}

  \node[yellow-bg,fit=(p1),name=c2]{};

 \begin{pgfonlayer}{foreground}
    \node at ($(dummy.east)+(4mm,0mm)$)  [yellow-move-place,name=dummy] {$d.dd$};
    \node at ($(dummy.west)+(0mm,4mm)$)  [white-move-place,name=p1] {$1.87$};
    \node at ($(dummy.west)+(0mm,-4mm)$)  [orange-move-place,name=p2] {$2.87$};
  \end{pgfonlayer}

  \node[yellow-bg,fit=(p1) (p2),name=c3]{};

 \begin{pgfonlayer}{foreground}
    \node at ($(dummy.east)+(4mm,0mm)$)  [yellow-move-place,name=dummy] {$d.dd$};
    \node at ($(dummy.west)+(0mm,4mm)$) [white-move-place,name=p1] {$2.90$};
    \node at ($(dummy.west)+(0mm,-4mm)$)  [blue-move-place,name=p2] {$8.90$};
  \end{pgfonlayer}

  \node[yellow-bg,fit=(p1) (p2),name=c4]{};

 \begin{pgfonlayer}{foreground}
   \node at ($(dummy.east)+(4mm,0mm)$)  [yellow-move-place,name=dummy] {$d.dd$};
    \node at ($(dummy.west)+(0mm,8mm)$) [red-move-place,name=p1] {$2.92$};
    \node at (dummy.west)  [green-move-place,name=p2] {$4.92$};
    \node at ($(dummy.west)+(0mm,-8mm)$) [orange-move-place,name=p3] {$4.92$};
  \end{pgfonlayer}

  \node[yellow-bg,fit=(p1) (p2) (p3),name=c5]{};

 \begin{pgfonlayer}{foreground}
   \node at ($(dummy.east)+(4mm,0mm)$)  [yellow-move-place,name=dummy] {$d.dd$};
     \node at (dummy.west)  [orange-move-place,name=p1] {$2.93$};
   \end{pgfonlayer}

  \node[yellow-bg,fit=(p1),name=c6]{};

\begin{pgfonlayer}{foreground}
   \node at ($(dummy.east)+(4mm,0mm)$)  [yellow-move-place,name=dummy] {$d.dd$};
     \node at (dummy.west)  [red-move-place,name=p1] {$1.96$};
   \end{pgfonlayer}

  \node[yellow-bg,fit=(p1),name=c7]{};

 \begin{pgfonlayer}{foreground}
    \node at ($(dummy.east)+(4mm,0mm)$) [yellow-move-place,name=dummy] {$d.dd$};
    \end{pgfonlayer}

  \node[yellow-bg,fit=(dummy),name=c75]{};

  \begin{pgfonlayer}{background}
    \node[black-bg,fit= (c1) (c2) (c3) (c4) (c5) (c6) (c7) (c75)]{};
  \end{pgfonlayer}

\end{scope}

\end{tikzpicture}
\caption{Detailed timed transitions (cont.).}
\label{detailed:cont:fig}
\end{figure}
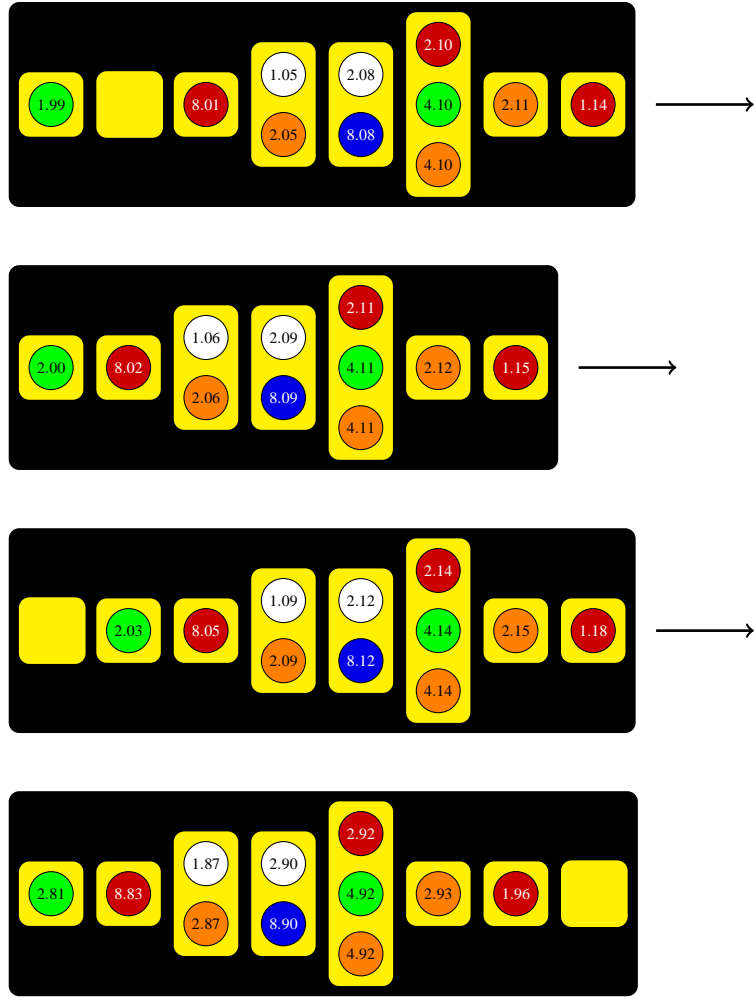

In the second step, the amount of delay is $0.02$
which is exactly the amount needed to allow the tokens that
currently have the highest fractional parts to become integers.
These tokens are the ones with ages $0.98$ and $1.98$ in 
\tikz{\node[white-ball]{};}
resp.\
\tikz{\node[orange-ball]{};}.
Their new ages are $1.00$ resp.\ $2.00$.
In the last step, all tokens have small fractional parts.
We let time pass sufficiently much ($0.78$ time units) so that the tokens
will all have high fractional parts.
Every computation of a {\sc Ptpn} can be transformed into an equivalent one 
(w.r.t. reachability and cost) where
all timed transitions are detailed, by replacing  long timed transitions with
several detailed shorter ones where necessary. 
Thus we may assume w.l.o.g.
that timed transitions are detailed.

\paragraph{Detailed Computations in $\delta$-form.}
In \cite{abdulla2011computing}, we show the following result.
For any computation $\comp$  
starting from an initial marking
$\initmarking$ (defined by a \emph{initial} place $\initp$),
and reaching a give set 
$\marking_{\finalp}$ of final markings 
(defined by a \emph{final} place $\finalp$),
and for each $\delta:0<\delta<0.2$,
there is a detailed computation $\comp'$ in $\delta$-form
where 
(i) $\comp'$ starts from the same initial marking as $\comp$,
(ii) $\comp'$ is in $\delta$-from,
(iii) $\comp'$ reaches  $\marking_{\finalp}$,
 and
(iv) if $\comp$ is detailed then $\comp'$ is detailed.
This means that, to solve the Cost-Threshold and Cost-Optimality
problems, it is sufficient to consider 
detailed computations in $\delta$-form.

\begin{figure}
\center

\begin{tikzpicture}[]

  \node[name=dummy]   {};
  
  \node[red-move-place,name=p1] at (dummy)  {$0.00$};
  \begin{pgfonlayer}{background}
    \node[black-bg,fit=(p1) ]{};
  \end{pgfonlayer}

  \node[arrow-node,name=arrow] at ($(p1.east)+(2mm,0mm)$) {$\longrightarrow$};
  \node[movelabel] at ($(arrow.center)+(0mm,3.5mm)$) {$1.01$};
  \node[moveprice] at ($(arrow.center)+(0mm,-3.5mm)$) {$3.03$};

  \node[red-move-place,name=p1] at ($(arrow.east)+(2mm,0)$) {$1.01$};
  \begin{pgfonlayer}{background}
    \node[black-bg,fit=(p1) ]{};
  \end{pgfonlayer}

  \node[arrow-node,name=arrow] at ($(p1.east)+(2mm,0mm)$) {$\longrightarrow$};
  \node[movelabel] at ($(arrow.center)+(0mm,3.5mm)$) {$t_1$};
  \node[moveprice] at ($(arrow.center)+(0mm,-3.5mm)$) {$2$};

  \node[white-move-place,name=p1] at ($(arrow.east)+(2mm,0)$) {$1.99$};
  \node[blue-move-place,name=p2] at ($(p1.east)+(1mm,0)$) {$3.01$};
  \begin{pgfonlayer}{background}
    \node[black-bg,fit=(p1) (p2)]{};
  \end{pgfonlayer}

  \node[arrow-node,name=arrow] at ($(p2.east)+(2mm,0mm)$) {$\longrightarrow$};
  \node[movelabel] at ($(arrow.center)+(0mm,3.5mm)$) {$0.02$};
  \node[moveprice] at ($(arrow.center)+(0mm,-3.5mm)$) {$0.02$};

  \node[white-move-place,name=p1] at ($(arrow.east)+(2mm,0)$) {$2.01$};
  \node[blue-move-place,name=p2] at ($(p1.east)+(1mm,0)$) {$3.03$};
  \begin{pgfonlayer}{background}
    \node[black-bg,fit=(p1) (p2)]{};
  \end{pgfonlayer}

  \node[arrow-node,name=arrow] at ($(p2.east)+(2mm,0mm)$) {$\longrightarrow$};
  \node[movelabel] at ($(arrow.center)+(0mm,3.5mm)$) {$t_2$};
  \node[moveprice] at ($(arrow.center)+(0mm,-3.5mm)$) {$4$};

  \node[green-move-place,name=p1] at ($(arrow.east)+(2mm,0)$) {$4.00$};
  \node[blue-move-place,name=p2] at ($(p1.east)+(1mm,0)$) {$3.03$};
  \begin{pgfonlayer}{background}
    \node[black-bg,fit=(p1) (p2)]{};
  \end{pgfonlayer}

  \node[arrow-node,name=arrow] at ($(p2.east)+(2mm,0mm)$) {$\longrightarrow$};
  \node[movelabel] at ($(arrow.center)+(0mm,3.5mm)$) {$0.99$};
  \node[moveprice] at ($(arrow.center)+(0mm,-3.5mm)$) {$1.98$};

  \node[name=dummy] at ($(dummy)+(0mm,-20mm)$)  {};
  \node[green-move-place,name=p1] at (dummy) {$4.99$};
  \node[blue-move-place,name=p2] at ($(p1.east)+(1mm,0)$) {$4.02$};
  \begin{pgfonlayer}{background}
    \node[black-bg,fit=(p1) (p2)]{};
  \end{pgfonlayer}

  \node[arrow-node,name=arrow] at ($(p2.east)+(2mm,0mm)$) {$\longrightarrow$};
  \node[movelabel] at ($(arrow.center)+(0mm,3.5mm)$) {$t_4$};
  \node[moveprice] at ($(arrow.center)+(0mm,-3.5mm)$) {$0$};

  \node[red-move-place,name=p1] at ($(arrow.east)+(2mm,0)$) {$2.00$};
  \node[blue-move-place,name=p2] at ($(p1.east)+(1mm,0)$) {$4.02$};
  \begin{pgfonlayer}{background}
    \node[black-bg,fit=(p1) (p2)]{};
  \end{pgfonlayer}

  \node[arrow-node,name=arrow] at ($(p2.east)+(2mm,0mm)$) {$\longrightarrow$};
  \node[movelabel] at ($(arrow.center)+(0mm,3.5mm)$) {$t_1$};
  \node[moveprice] at ($(arrow.center)+(0mm,-3.5mm)$) {$2$};

  \node[white-move-place,name=p1] at ($(arrow.east)+(2mm,0)$) {$0.99$};
  \node[blue-move-place,name=p2] at ($(p1.east)+(2mm,0)$) {$3.00$};
  \node[blue-move-place,name=p3] at ($(p2.east)+(1mm,0)$) {$4.02$};
  \begin{pgfonlayer}{background}
    \node[black-bg,fit=(p1) (p2) (p3)]{};
  \end{pgfonlayer}

  \node[arrow-node,name=arrow] at ($(p3.east)+(2mm,0mm)$) {$\longrightarrow$};
  \node[movelabel] at ($(arrow.center)+(0mm,3.5mm)$) {$0.02$};
  \node[moveprice] at ($(arrow.center)+(0mm,-3.5mm)$) {$0.02$};

  \node[white-move-place,name=p1] at ($(arrow.east)+(2mm,0)$) {$1.01$};
  \node[blue-move-place,name=p2] at ($(p1.east)+(2mm,0)$) {$3.02$};
  \node[blue-move-place,name=p3] at ($(p2.east)+(1mm,0)$) {$4.04$};
  \begin{pgfonlayer}{background}
    \node[black-bg,fit=(p1) (p2) (p3)]{};
  \end{pgfonlayer}

  \node[arrow-node,name=arrow] at ($(p3.east)+(2mm,0mm)$) {$\longrightarrow$};
  \node[movelabel] at ($(arrow.center)+(0mm,3.5mm)$) {$t_2$};
  \node[moveprice] at ($(arrow.center)+(0mm,-3.5mm)$) {$4$};

  \node[name=dummy] at ($(dummy)+(0mm,-20mm)$)  {};
  \node[green-move-place,name=p1] at (dummy) {$4.00$};
  \node[blue-move-place,name=p2] at ($(p1.east)+(1mm,0)$) {$3.02$};
  \node[blue-move-place,name=p3] at ($(p2.east)+(1mm,0)$) {$4.04$};
  \begin{pgfonlayer}{background}
    \node[black-bg,fit=(p1) (p2) (p3)]{};
  \end{pgfonlayer}

  \node[arrow-node,name=arrow] at ($(p3.east)+(2mm,0mm)$) {$\longrightarrow$};
  \node[movelabel] at ($(arrow.center)+(0mm,3.5mm)$) {$t_3$};
  \node[moveprice] at ($(arrow.center)+(0mm,-3.5mm)$) {$3$};

  \node[green-move-place,name=p1] at ($(arrow.east)+(2mm,0)$) {$4.00$};
  \node[orange-move-place,name=p2] at ($(p1.east)+(2mm,0)$) {$1.99$};
  \node[blue-move-place,name=p3] at ($(p2.east)+(1mm,0)$) {$4.04$};
  \begin{pgfonlayer}{background}
    \node[black-bg,fit=(p1) (p2) (p3)]{};
  \end{pgfonlayer}

  \node[arrow-node,name=arrow] at ($(p3.east)+(2mm,0mm)$) {$\longrightarrow$};
  \node[movelabel] at ($(arrow.center)+(0mm,3.5mm)$) {$0.02$};
  \node[moveprice] at ($(arrow.center)+(0mm,-3.5mm)$) {$0.04$};

  \node[green-move-place,name=p1] at ($(arrow.east)+(2mm,0)$) {$4.02$};
  \node[orange-move-place,name=p2] at ($(p1.east)+(2mm,0)$) {$2.01$};
  \node[blue-move-place,name=p3] at ($(p2.east)+(1mm,0)$) {$4.06$};
  \begin{pgfonlayer}{background}
    \node[black-bg,fit=(p1) (p2) (p3)]{};
  \end{pgfonlayer}

  \node[arrow-node,name=arrow] at ($(p3.east)+(2mm,0mm)$) {$\longrightarrow$};
  \node[movelabel] at ($(arrow.center)+(0mm,3.5mm)$) {$t_5$};
  \node[moveprice] at ($(arrow.center)+(0mm,-3.5mm)$) {$0$};

  \node[red-move-place,name=p1] at ($(arrow.east)+(2mm,0)$) {$1.01$};
  \node[blue-move-place,name=p2] at ($(p1.east)+(1mm,0)$) {$4.06$};
  \begin{pgfonlayer}{background}
    \node[black-bg,fit=(p1) (p2)]{};
  \end{pgfonlayer}

\end{tikzpicture}

\label{detailed:comp:delta:form:fig}
\caption{A detailed computation in $\delta$-form.}
\end{figure}
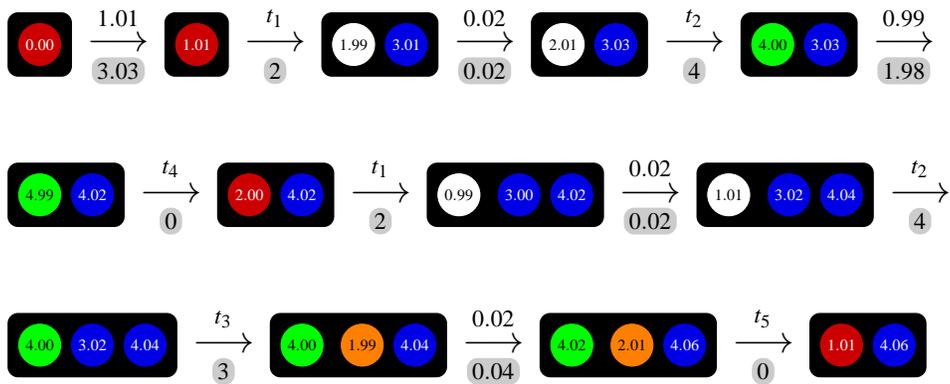

Figure~\ref{detailed:comp:delta:form:fig} shows a detailed
computation in $\delta$-form for the {\sc Ptpn}
of Figure~\ref{ptpn:figure}.

\section{Regions}
\label{regions:section}
In this section, we introduce a symbolic encoding for infinite sets of markings.
The encoding is a variant of the classical
notion of {\it regions}  \cite{AD:timedautomata}.
The main difference is that 
we here need to deal with an unbounded number of clocks.
It is an adaptation of the encoding introduced in 
\cite{abdulla2003model}.
More precisely, we change the encoding of \cite{abdulla2003model}
so that we can now deal with markings in $\delta$-form.
First, we give the definition of regions, and then we show how to
simulate timed and discrete transitions on regions.
For each type of transition, we define the 
cost of firing the transition from the region.

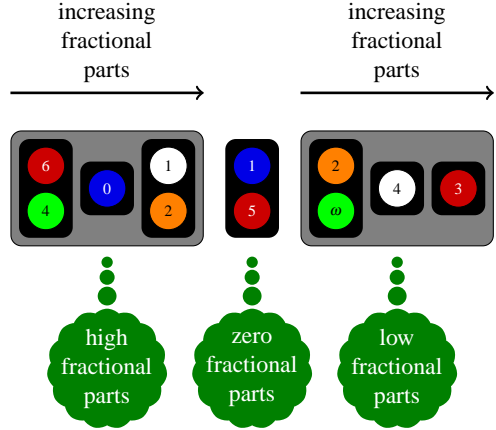
\begin{figure}
\center
\begin{tikzpicture}[]

 \begin{pgfonlayer}{foreground}
    \node   [black-move-place,name=dummy] {$d$};
    \node at ($(dummy.west)+(0mm,3mm)$)  [red-move-place,name=p1] {$6$};
    \node at ($(dummy.west)+(0mm,-3mm)$)  [green-move-place,name=p2] {$4$};
  \end{pgfonlayer}

  \node[regionnode,fit=(p1) (p2),name=c1]{};

 \begin{pgfonlayer}{foreground}
    \node at ($(dummy.east)+(3mm,0mm)$)  [black-move-place,name=dummy] {$d$};
    \node at ($(dummy.west)+(0mm,0mm)$) [blue-move-place,name=p1] {$0$};
  \end{pgfonlayer}

  \node[regionnode,fit=(p1) ,name=c2]{};

 \begin{pgfonlayer}{foreground}
    \node at ($(dummy.east)+(3mm,0mm)$)   [black-move-place,name=dummy] {$d$};
    \node at ($(dummy.west)+(0mm,3mm)$)  [white-move-place,name=p1] {$1$};
    \node at ($(dummy.west)+(0mm,-3mm)$)  [orange-move-place,name=p2] {$2$};
  \end{pgfonlayer}

  \node[regionnode,fit=(p1) (p2),name=c3]{};

\begin{pgfonlayer}{background}
    \node [fit=(c1) (c2) (c3),Regionnode,name=r1] {};
  \end{pgfonlayer}

\draw[line width=1pt,->] ($(r1.north west)+(0mm,5mm)$) -- node[above] 
{\begin{minipage}{0.1\textwidth}\begin{center}\footnotesize increasing\\ fractional\\ parts\end{center}\end{minipage}} 
($(r1.north east)+(0mm,5mm)$);

\node[
line width=1pt,
fill=green!50!black, text=white,
cloud,
cloud puffs=15, 
cloud puff arc=90,
%fill=red!70!black,
minimum width=1.6cm,
minimum height=1.6cm,
name=cloud0]
at ($(c2)+(0mm,-24mm)$)
{};

\node[] at (cloud0){\footnotesize\begin{minipage}{0.1\textwidth}\begin{center}\white{high\\fractional \\parts}\end{center}\end{minipage}};

\node[circle,fill=green!50!black,inner sep =0mm,minimum size =2.5 mm] at ($(cloud0.north)+(0mm,1.5mm)$){};
\node[circle,fill=green!50!black,inner sep =0mm,minimum size =2.0 mm] at ($(cloud0.north)+(-0mm,4mm)$){};
\node[circle,fill=green!50!black,inner sep =0mm,minimum size =1.5 mm] at ($(cloud0.north)+(0mm,6mm)$){};

  \begin{pgfonlayer}{foreground}
    \node at ($(dummy.east)+(6mm,0mm)$)   [black-move-place,name=dummy] {$d$};
    \node at ($(dummy.west)+(0mm,3mm)$)  [blue-move-place,name=p1] {$1$};
    \node at ($(dummy.west)+(0mm,-3mm)$)  [red-move-place,name=p2] {$5$};
  \end{pgfonlayer}

  \node[regionnode,fit=(p1) (p2),name=c3]{};

\node[
line width=1pt,
fill=green!50!black, text=white,
cloud,
cloud puffs=15, 
cloud puff arc=90,
%fill=red!70!black,
minimum width=1.6cm,
minimum height=1.6cm,
name=cloud0]
at ($(c3)+(0mm,-24mm)$)
{};

\node[] at (cloud0){\footnotesize\begin{minipage}{0.1\textwidth}\begin{center}\white{zero\\fractional \\parts}\end{center}\end{minipage}};

\node[circle,fill=green!50!black,inner sep =0mm,minimum size =2.5 mm] at ($(cloud0.north)+(0mm,1.5mm)$){};
\node[circle,fill=green!50!black,inner sep =0mm,minimum size =2.0 mm] at ($(cloud0.north)+(-0mm,4mm)$){};
\node[circle,fill=green!50!black,inner sep =0mm,minimum size =1.5 mm] at ($(cloud0.north)+(0mm,6mm)$){};

 \begin{pgfonlayer}{foreground}
    \node at ($(dummy.east)+(6mm,0mm)$)    [black-move-place,name=dummy] {$d$};
    \node at ($(dummy.west)+(0mm,3mm)$)  [orange-move-place,name=p1] {$2$};
    \node at ($(dummy.west)+(0mm,-3mm)$)  [green-move-place,name=p2] {$\omega$};
  \end{pgfonlayer}

  \node[regionnode,fit=(p1) (p2),name=c1]{};

 \begin{pgfonlayer}{foreground}
    \node at ($(dummy.east)+(3mm,0mm)$)  [black-move-place,name=dummy] {$d$};
    \node at ($(dummy.west)+(0mm,0mm)$) [white-move-place,name=p1] {$4$};
  \end{pgfonlayer}

  \node[regionnode,fit=(p1) ,name=c2]{};

 \begin{pgfonlayer}{foreground}
    \node at ($(dummy.east)+(3mm,0mm)$)   [black-move-place,name=dummy] {$d$};
    \node at ($(dummy.west)+(0mm,0mm)$)  [red-move-place,name=p1] {$3$};
  \end{pgfonlayer}

  \node[regionnode,fit=(p1),name=c3]{};

\begin{pgfonlayer}{background}
    \node [fit=(c1) (c2) (c3),Regionnode,name=r1] {};
  \end{pgfonlayer}

\node[
line width=1pt,
fill=green!50!black, text=white,
cloud,
cloud puffs=15, 
cloud puff arc=90,
%fill=red!70!black,
minimum width=1.6cm,
minimum height=1.6cm,
name=cloud0]
at ($(c2)+(0mm,-24mm)$)
{};

\node[] at (cloud0){\footnotesize\begin{minipage}{0.1\textwidth}\begin{center}\white{low\\fractional \\parts}\end{center}\end{minipage}};

\node[circle,fill=green!50!black,inner sep =0mm,minimum size =2.5 mm] at ($(cloud0.north)+(0mm,1.5mm)$){};
\node[circle,fill=green!50!black,inner sep =0mm,minimum size =2.0 mm] at ($(cloud0.north)+(-0mm,4mm)$){};
\node[circle,fill=green!50!black,inner sep =0mm,minimum size =1.5 mm] at ($(cloud0.north)+(0mm,6mm)$){};

\draw[line width=1pt,->] ($(r1.north west)+(0mm,5mm)$) -- node[above] 
{\begin{minipage}{0.1\textwidth}\begin{center}\footnotesize increasing\\ fractional\\ parts\end{center}\end{minipage}} 
($(r1.north east)+(0mm,5mm)$);

\end{tikzpicture}

\caption{A region $\region$.}
\label{region:fig}
\end{figure}

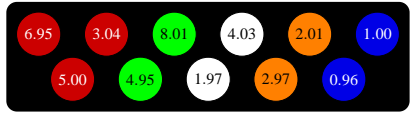
\begin{figure}
\center

\begin{tikzpicture}[]
 \node[red-move-place,name=n1] {$6.95$};
  \node at ($(n1.east)+(3mm,0mm)$) [red-move-place,name=n3] {$3.04$};
  \node at ($(n3.east)+(3mm,0mm)$) [green-move-place,name=n5] {$8.01$};
  \node at ($(n5.east)+(3mm,0mm)$) [white-move-place,name=n7] {$4.03$};
  \node at ($(n7.east)+(3mm,0mm)$) [orange-move-place,name=n9] {$2.01$};
  \node at ($(n9.east)+(3mm,0mm)$) [blue-move-place,name=n11] {$1.00$};
  \node at ($(n1.center) !.5! (n3.center)+(0mm,-6mm)$) [red-move-place,anchor=center,name=n2] {$5.00$};
  \node at ($(n2.east)+(3mm,0mm)$) [green-move-place,name=n4] {$4.95$};
  \node at ($(n4.east)+(3mm,0mm)$) [white-move-place,name=n6] {$1.97$};
  \node at ($(n6.east)+(3mm,0mm)$) [orange-move-place,name=n8] {$2.97$};
  \node at ($(n8.east)+(3mm,0mm)$) [blue-move-place,name=n10] {$0.96$};
\begin{pgfonlayer}{background}
\node[black-bg,fit=(n1) (n2) (n3) (n4) (n5) (n6) (n7) (n8) (n9) (n10) (n11)]{};
\end{pgfonlayer}

\end{tikzpicture}

\caption{A marking $\marking$ satisfying the region of Figure~\ref{region:fig}.}
\label{region:marking:fig}
\end{figure}

\paragraph{Regions.}
A region characterizes a set of marking in $\delta$-form for some
$\delta:0<\delta<0.2$.
An example of (our notion of) a region $\region$
is shown in Figure~\ref{region:fig}.
The region consists of three
parts, referred to as $H$ (for high),
$Z$ (for zero), and
$L$ (for low).
The part $H$ is a word of multisets.
Each element in a multiset  
is a colored ball with a natural number,
representing one token.
The color defines the place in which the token resides, while
the number defines the integer part of the age of the token.
Furthermore, tokens whose ages are larger than $\maxval+1$ are all
represented by one element $\omega$
(ages $>\maxval$ cannot be distinguished by the transitions
of the {\sc Ptpn}).
The ordering of the multisets reflects the ordering of the factional
parts of the corresponding tokens:
elements belonging to the same multiset represent tokens with
identical fractional parts, and elements
in successive multisets represent tokens with increasing
fractional parts.
The part $Z$ consists of one multiset, and represents the tokens
with zero fractional parts.
Finally, the part $L$ consists of a word of multisets.
It has a similar interpretation to $H$, except that it
 represents tokens with low fractional parts.
Figure~\ref{region:marking:fig} shows a marking $\marking$ 
(of the Petri net of Figure~\ref{ptpn:figure}) satisfying the 
region $\region$ of Figure~\ref{region:fig} as follows:
\begin{itemize}
\item
The left-most multiset in $H$ contains 
a red ball with value $6$ and a green ball with value $4$.
They represent the token with age $6.95$ in 
the place
\tikz{\node[red-ball]{};}, and the token with age $4.95$ in 
\tikz{\node[green-ball]{};}.
The fractional parts of the two tokens are equal
($0.95$) and high.
\item
The next multiset contains a blue ball with value
$0$.
It represents the token with age $0.96$ in 
\tikz{\node[blue-ball]{};}.
The fractional part of the token ($0.96$) is high  and is
larger than the fractional parts
of the tokens in the previous multiset.
\item
The right-most multiset in $H$ contains 
a white ball with value $1$ and an orange ball with value $2$.
They represent the token with ages $1.97$ in 
the place
\tikz{\node[white-ball]{};}, and the token with age $2.97$ in 
\tikz{\node[orange-ball]{};}.
The fractional parts of the two tokens are equal
($0.97$).
The fractional parts of these tokens ($0.97$) are high and are
larger than the fractional part
of the token in the previous multiset.

\item
The part $Z$ consists of a single multiset.
It contains 
a blue ball with value $1$ and a red ball with value $5$.
They represent the token with age $1.00$ in 
the place
\tikz{\node[blue-ball]{};}, and the token with age $5$ in 
\tikz{\node[red-ball]{};}.
The fractional parts of the two tokens are zero. 
\item
The left-most multiset in $L$ contains 
an orange ball with value $2$ and a green ball with value $\omega$.
They represent the token with age $2.01$ in 
the place
\tikz{\node[orange-ball]{};}, and the token with age $8.01$ in 
\tikz{\node[green-ball]{};}.
The fractional parts of the two tokens are equal
($0.01$) and low.
The age of the token in 
\tikz{\node[green-ball]{};} is
$8.01\geq\maxval+1$ which means that it is represented by 
$\omega$ in $\region$.
\item
The next multiset contains a white ball with value
$4$.
It represents that token with age $4.03$ in 
\tikz{\node[white-ball]{};}.
The fractional part of the token ($0.03$) is low  and is
larger than the fractional parts
of the tokens in the previous multiset.
\item
The next multiset contains a red ball with value
$3$.
It represents that token with age $3.04$ in 
\tikz{\node[red-ball]{};}.
The fractional part of the token ($0.04$) is low  and is
larger than the fractional part
of the token in the previous multiset.

\end{itemize}

We use $\denotationof\region$ to denote the set of markings satisfying $\region$.

\paragraph{Timed Transitions.}
We will
describe how to encode the effect of detailed timed transitions
on regions.
To do that, we define 4 different types of transitions on regions.
\begin{description}
\item[Type I]
This simulates a small delay  where 
the tokens of integer age now have a positive fractional part,
but no tokens reach an integer age.
An example of such a transition is shown in Figure~\ref{type:1:fig}.
Here, the delay is $0.01$ which is not sufficient to make the tokens
with the highest fractional parts
(the token with age $1.97$ in
\tikz{\node[white-ball]{};}, and the token with age $2.97$ in
\tikz{\node[orange-ball]{};})
to become integers.
Notice that the tokens with zero fractional parts
(the token with age $1.00$ in
\tikz{\node[blue-ball]{};}, and the token with age $5.00$ in
\tikz{\node[red-ball]{};})
will now have have low fractional parts (in fact,
they will have the smallest fractional parts, namely $0.01$,
among all tokens
in the marking).
At the region level, the two elements in $Z$ will move to $L$, 
forming the left-most multiset in $L$ (reflecting
the fact that they have the lowest fractional parts).
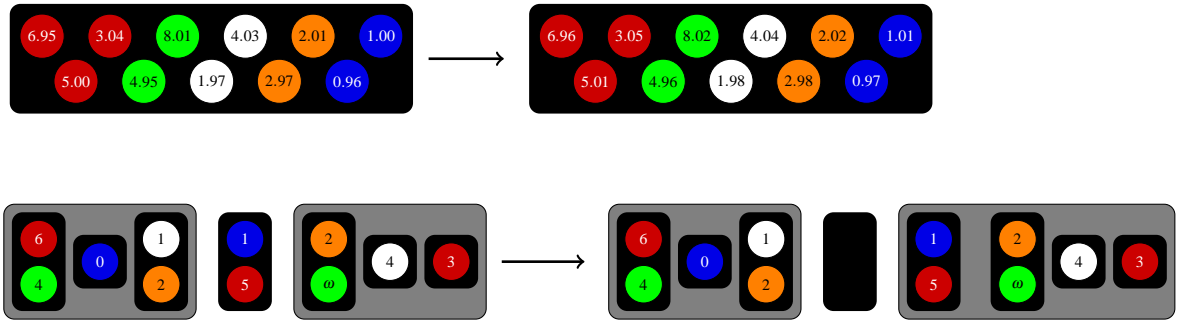
\begin{figure}
\begin{tikzpicture}[]
  \node[name=dummy] {};
  \node at (dummy) [red-move-place,name=n1] {$6.95$};
  \node at ($(n1.east)+(3mm,0mm)$) [red-move-place,name=n3] {$3.04$};
  \node at ($(n3.east)+(3mm,0mm)$) [green-move-place,name=n5] {$8.01$};
  \node at ($(n5.east)+(3mm,0mm)$) [white-move-place,name=n7] {$4.03$};
  \node at ($(n7.east)+(3mm,0mm)$) [orange-move-place,name=n9] {$2.01$};
  \node at ($(n9.east)+(3mm,0mm)$) [blue-move-place,name=n11] {$1.00$};
  \node at ($(n1.center) !.5! (n3.center)+(0mm,-6mm)$) [red-move-place,anchor=center,name=n2] {$5.00$};
  \node at ($(n2.east)+(3mm,0mm)$) [green-move-place,name=n4] {$4.95$};
  \node at ($(n4.east)+(3mm,0mm)$) [white-move-place,name=n6] {$1.97$};
  \node at ($(n6.east)+(3mm,0mm)$) [orange-move-place,name=n8] {$2.97$};
  \node at ($(n8.east)+(3mm,0mm)$) [blue-move-place,name=n10] {$0.96$};
\begin{pgfonlayer}{background}
\node[black-bg,fit=(n1) (n2) (n3) (n4) (n5) (n6) (n7) (n8) (n9) (n10) (n11),name=r]{};
\end{pgfonlayer}

\draw[line width=1pt,->] ($(r.east)+(2mm,0mm)$) -- ($(r.east)+(12mm,0mm)$);

  \node at ($(n11.east)+(18mm,0mm)$) [red-move-place,name=n1] {$6.96$};
  \node at ($(n1.east)+(3mm,0mm)$) [red-move-place,name=n3] {$3.05$};
  \node at ($(n3.east)+(3mm,0mm)$) [green-move-place,name=n5] {$8.02$};
  \node at ($(n5.east)+(3mm,0mm)$) [white-move-place,name=n7] {$4.04$};
  \node at ($(n7.east)+(3mm,0mm)$) [orange-move-place,name=n9] {$2.02$};
  \node at ($(n9.east)+(3mm,0mm)$) [blue-move-place,name=n11] {$1.01$};
  \node at ($(n1.center) !.5! (n3.center)+(0mm,-6mm)$) [red-move-place,anchor=center,name=n2] {$5.01$};
  \node at ($(n2.east)+(3mm,0mm)$) [green-move-place,name=n4] {$4.96$};
  \node at ($(n4.east)+(3mm,0mm)$) [white-move-place,name=n6] {$1.98$};
  \node at ($(n6.east)+(3mm,0mm)$) [orange-move-place,name=n8] {$2.98$};
  \node at ($(n8.east)+(3mm,0mm)$) [blue-move-place,name=n10] {$0.97$};
\begin{pgfonlayer}{background}
\node[black-bg,fit=(n1) (n2) (n3) (n4) (n5) (n6) (n7) (n8) (n9) (n10) (n11)]{};
\end{pgfonlayer}

%%%%%%%%%%%%%%%%%%  Regions %%%%%%

 \begin{pgfonlayer}{foreground}
\node at ($(dummy)+(0mm,-30mm)$)  [black-move-place,name=dummy] {$d$};
    \node at ($(dummy.west)+(0mm,3mm)$)  [red-move-place,name=p1] {$6$};
    \node at ($(dummy.west)+(0mm,-3mm)$)  [green-move-place,name=p2] {$4$};
  \end{pgfonlayer}

  \node[regionnode,fit=(p1) (p2),name=c1]{};

 \begin{pgfonlayer}{foreground}
    \node at ($(dummy.east)+(3mm,0mm)$)  [black-move-place,name=dummy] {$d$};
    \node at ($(dummy.west)+(0mm,0mm)$) [blue-move-place,name=p1] {$0$};
  \end{pgfonlayer}

  \node[regionnode,fit=(p1) ,name=c2]{};

 \begin{pgfonlayer}{foreground}
    \node at ($(dummy.east)+(3mm,0mm)$)   [black-move-place,name=dummy] {$d$};
    \node at ($(dummy.west)+(0mm,3mm)$)  [white-move-place,name=p1] {$1$};
    \node at ($(dummy.west)+(0mm,-3mm)$)  [orange-move-place,name=p2] {$2$};
  \end{pgfonlayer}

  \node[regionnode,fit=(p1) (p2),name=c3]{};

\begin{pgfonlayer}{background}
    \node [fit=(c1) (c2) (c3),Regionnode,name=r1] {};
  \end{pgfonlayer}

  \begin{pgfonlayer}{foreground}
    \node at ($(dummy.east)+(6mm,0mm)$)   [black-move-place,name=dummy] {$d$};
    \node at ($(dummy.west)+(0mm,3mm)$)  [blue-move-place,name=p1] {$1$};
    \node at ($(dummy.west)+(0mm,-3mm)$)  [red-move-place,name=p2] {$5$};
  \end{pgfonlayer}

  \node[regionnode,fit=(p1) (p2),name=c3]{};

 \begin{pgfonlayer}{foreground}
    \node at ($(dummy.east)+(6mm,0mm)$)    [black-move-place,name=dummy] {$d$};
    \node at ($(dummy.west)+(0mm,3mm)$)  [orange-move-place,name=p1] {$2$};
    \node at ($(dummy.west)+(0mm,-3mm)$)  [green-move-place,name=p2] {$\omega$};
  \end{pgfonlayer}

  \node[regionnode,fit=(p1) (p2),name=c1]{};

 \begin{pgfonlayer}{foreground}
    \node at ($(dummy.east)+(3mm,0mm)$)  [black-move-place,name=dummy] {$d$};
    \node at ($(dummy.west)+(0mm,0mm)$) [white-move-place,name=p1] {$4$};
  \end{pgfonlayer}

  \node[regionnode,fit=(p1) ,name=c2]{};

 \begin{pgfonlayer}{foreground}
    \node at ($(dummy.east)+(3mm,0mm)$)   [black-move-place,name=dummy] {$d$};
    \node at ($(dummy.west)+(0mm,0mm)$)  [red-move-place,name=p1] {$3$};
  \end{pgfonlayer}

  \node[regionnode,fit=(p1),name=c3]{};

\begin{pgfonlayer}{background}
    \node [fit=(c1) (c2) (c3),Regionnode,name=r1] {};
  \end{pgfonlayer}

\draw[line width=1pt,->] ($(r1.east)+(2mm,0mm)$) -- ($(r1.east)+(13mm,0mm)$);

 \begin{pgfonlayer}{foreground}
\node at ($(c3)+(23mm,0mm)$)  [black-move-place,name=dummy] {$d$};
    \node at ($(dummy.west)+(0mm,3mm)$)  [red-move-place,name=p1] {$6$};
    \node at ($(dummy.west)+(0mm,-3mm)$)  [green-move-place,name=p2] {$4$};
  \end{pgfonlayer}

  \node[regionnode,fit=(p1) (p2),name=c1]{};

 \begin{pgfonlayer}{foreground}
    \node at ($(dummy.east)+(3mm,0mm)$)  [black-move-place,name=dummy] {$d$};
    \node at ($(dummy.west)+(0mm,0mm)$) [blue-move-place,name=p1] {$0$};
  \end{pgfonlayer}

  \node[regionnode,fit=(p1) ,name=c2]{};

\begin{pgfonlayer}{foreground}
    \node at ($(dummy.east)+(3mm,0mm)$)   [black-move-place,name=dummy] {$d$};
    \node at ($(dummy.west)+(0mm,3mm)$)  [white-move-place,name=p1] {$1$};
    \node at ($(dummy.west)+(0mm,-3mm)$)  [orange-move-place,name=p2] {$2$};
  \end{pgfonlayer}

  \node[regionnode,fit=(p1) (p2),name=c3]{};

  \begin{pgfonlayer}{background}
    \node [fit= (c1) (c2) (c3),Regionnode,name=r1] {};
  \end{pgfonlayer}

\begin{pgfonlayer}{foreground}
    \node at ($(dummy.east)+(6mm,0mm)$)   [black-move-place,name=dummy] {$d$};
    \node at ($(dummy.west)+(0mm,3mm)$)  [black-move-place,name=p1] {$1$};
    \node at ($(dummy.west)+(0mm,-3mm)$)  [black-move-place,name=p2] {$5$};
  \end{pgfonlayer}

  \node[regionnode,fit=(p1) (p2),name=c4]{};

  \begin{pgfonlayer}{foreground}
    \node at ($(dummy.east)+(6mm,0mm)$)   [black-move-place,name=dummy] {$d$};
    \node at ($(dummy.west)+(0mm,3mm)$)  [blue-move-place,name=p1] {$1$};
    \node at ($(dummy.west)+(0mm,-3mm)$)  [red-move-place,name=p2] {$5$};
  \end{pgfonlayer}

  \node[regionnode,fit=(p1) (p2),name=c0]{};

 \begin{pgfonlayer}{foreground}
    \node at ($(dummy.east)+(6mm,0mm)$)    [black-move-place,name=dummy] {$d$};
    \node at ($(dummy.west)+(0mm,3mm)$)  [orange-move-place,name=p1] {$2$};
    \node at ($(dummy.west)+(0mm,-3mm)$)  [green-move-place,name=p2] {$\omega$};
  \end{pgfonlayer}

  \node[regionnode,fit=(p1) (p2),name=c1]{};

 \begin{pgfonlayer}{foreground}
    \node at ($(dummy.east)+(3mm,0mm)$)  [black-move-place,name=dummy] {$d$};
    \node at ($(dummy.west)+(0mm,0mm)$) [white-move-place,name=p1] {$4$};
  \end{pgfonlayer}

  \node[regionnode,fit=(p1) ,name=c2]{};

 \begin{pgfonlayer}{foreground}
    \node at ($(dummy.east)+(3mm,0mm)$)   [black-move-place,name=dummy] {$d$};
    \node at ($(dummy.west)+(0mm,0mm)$)  [red-move-place,name=p1] {$3$};
  \end{pgfonlayer}

  \node[regionnode,fit=(p1),name=c3]{};

\begin{pgfonlayer}{background}
    \node [fit= (c0) (c1) (c2) (c3),Regionnode,name=r1] {};
  \end{pgfonlayer}

\end{tikzpicture}

\caption{Type I Transition.}
\label{type:1:fig}
\end{figure}

\item[Type II Transition.]
This simulates a small delay in the case where
there were no tokens of integer age
and the tokens with the
highest fractional parts just reach the next integer age.
An example of such a transition is shown in Figure~\ref{type:2:fig}.
Here, the delay is $0.02$, which is sufficient to make the tokens
with the highest fractional parts
(the token with age $1.98$ in
\tikz{\node[white-ball]{};}, and the token with age $2.98$ in
\tikz{\node[orange-ball]{};})
to become integers, i.e., $2$ and $3$ respectively.
At the region level, the right-most multiset
in $H$ will move to $Z$,
and the value of each element in the multiset is incremented
by one  
to reflect the fact that the ages of the token moves to the next integer.
\begin{figure}

\begin{tikzpicture}[]
  \node[name=dummy] {};
  \node at (dummy) [red-move-place,name=n1] {$6.96$};
  \node at ($(n1.east)+(3mm,0mm)$) [red-move-place,name=n3] {$3.05$};
  \node at ($(n3.east)+(3mm,0mm)$) [green-move-place,name=n5] {$8.02$};
  \node at ($(n5.east)+(3mm,0mm)$) [white-move-place,name=n7] {$4.04$};
  \node at ($(n7.east)+(3mm,0mm)$) [orange-move-place,name=n9] {$2.02$};
  \node at ($(n9.east)+(3mm,0mm)$) [blue-move-place,name=n11] {$1.01$};
  \node at ($(n1.center) !.5! (n3.center)+(0mm,-6mm)$) [red-move-place,anchor=center,name=n2] {$5.01$};
  \node at ($(n2.east)+(3mm,0mm)$) [green-move-place,name=n4] {$4.96$};
  \node at ($(n4.east)+(3mm,0mm)$) [white-move-place,name=n6] {$1.98$};
  \node at ($(n6.east)+(3mm,0mm)$) [orange-move-place,name=n8] {$2.98$};
  \node at ($(n8.east)+(3mm,0mm)$) [blue-move-place,name=n10] {$0.97$};
\begin{pgfonlayer}{background}
\node[black-bg,fit=(n1) (n2) (n3) (n4) (n5) (n6) (n7) (n8) (n9) (n10) (n11),name=r]{};
\end{pgfonlayer}

\draw[line width=1pt,->] ($(r.east)+(2mm,0mm)$) -- ($(r.east)+(12mm,0mm)$);

  \node at ($(n11.east)+(18mm,0mm)$)  [red-move-place,name=n1] {$6.98$};
  \node at ($(n1.east)+(3mm,0mm)$) [red-move-place,name=n3] {$3.07$};
  \node at ($(n3.east)+(3mm,0mm)$) [green-move-place,name=n5] {$8.04$};
  \node at ($(n5.east)+(3mm,0mm)$) [white-move-place,name=n7] {$4.06$};
  \node at ($(n7.east)+(3mm,0mm)$) [orange-move-place,name=n9] {$2.04$};
  \node at ($(n9.east)+(3mm,0mm)$) [blue-move-place,name=n11] {$1.03$};
  \node at ($(n1.center) !.5! (n3.center)+(0mm,-6mm)$) [red-move-place,anchor=center,name=n2] {$5.03$};
  \node at ($(n2.east)+(3mm,0mm)$) [green-move-place,name=n4] {$4.98$};
  \node at ($(n4.east)+(3mm,0mm)$) [white-move-place,name=n6] {$2.00$};
  \node at ($(n6.east)+(3mm,0mm)$) [orange-move-place,name=n8] {$3.00$};
  \node at ($(n8.east)+(3mm,0mm)$) [blue-move-place,name=n10] {$0.99$};
\begin{pgfonlayer}{background}
\node[black-bg,fit=(n1) (n2) (n3) (n4) (n5) (n6) (n7) (n8) (n9) (n10) (n11),name=r]{};
\end{pgfonlayer}

%%%%%%%%%%%%%%%%%%  Regions %%%%%%

 \begin{pgfonlayer}{foreground}
\node at ($(dummy)+(0mm,-30mm)$)  [black-move-place,name=dummy] {$d$};
    \node at ($(dummy.west)+(0mm,3mm)$)  [red-move-place,name=p1] {$6$};
    \node at ($(dummy.west)+(0mm,-3mm)$)  [green-move-place,name=p2] {$4$};
  \end{pgfonlayer}

  \node[regionnode,fit=(p1) (p2),name=c1]{};

 \begin{pgfonlayer}{foreground}
    \node at ($(dummy.east)+(3mm,0mm)$)  [black-move-place,name=dummy] {$d$};
    \node at ($(dummy.west)+(0mm,0mm)$) [blue-move-place,name=p1] {$0$};
  \end{pgfonlayer}

  \node[regionnode,fit=(p1) ,name=c2]{};

\begin{pgfonlayer}{foreground}
    \node at ($(dummy.east)+(3mm,0mm)$)   [black-move-place,name=dummy] {$d$};
    \node at ($(dummy.west)+(0mm,3mm)$)  [white-move-place,name=p1] {$1$};
    \node at ($(dummy.west)+(0mm,-3mm)$)  [orange-move-place,name=p2] {$2$};
  \end{pgfonlayer}

  \node[regionnode,fit=(p1) (p2),name=c3]{};

  \begin{pgfonlayer}{background}
    \node [fit= (c1) (c2) (c3),Regionnode,name=r1] {};
  \end{pgfonlayer}

\begin{pgfonlayer}{foreground}
    \node at ($(dummy.east)+(6mm,0mm)$)   [black-move-place,name=dummy] {$d$};
    \node at ($(dummy.west)+(0mm,3mm)$)  [black-move-place,name=p1] {$1$};
    \node at ($(dummy.west)+(0mm,-3mm)$)  [black-move-place,name=p2] {$2$};
  \end{pgfonlayer}

  \node[regionnode,fit=(p1) (p2),name=c4]{};

  \begin{pgfonlayer}{foreground}
    \node at ($(dummy.east)+(6mm,0mm)$)   [black-move-place,name=dummy] {$d$};
    \node at ($(dummy.west)+(0mm,3mm)$)  [blue-move-place,name=p1] {$1$};
    \node at ($(dummy.west)+(0mm,-3mm)$)  [red-move-place,name=p2] {$5$};
  \end{pgfonlayer}

  \node[regionnode,fit=(p1) (p2),name=c0]{};

 \begin{pgfonlayer}{foreground}
    \node at ($(dummy.east)+(6mm,0mm)$)    [black-move-place,name=dummy] {$d$};
    \node at ($(dummy.west)+(0mm,3mm)$)  [orange-move-place,name=p1] {$2$};
    \node at ($(dummy.west)+(0mm,-3mm)$)  [green-move-place,name=p2] {$\omega$};
  \end{pgfonlayer}

  \node[regionnode,fit=(p1) (p2),name=c1]{};

 \begin{pgfonlayer}{foreground}
    \node at ($(dummy.east)+(3mm,0mm)$)  [black-move-place,name=dummy] {$d$};
    \node at ($(dummy.west)+(0mm,0mm)$) [white-move-place,name=p1] {$4$};
  \end{pgfonlayer}

  \node[regionnode,fit=(p1) ,name=c2]{};

 \begin{pgfonlayer}{foreground}
    \node at ($(dummy.east)+(3mm,0mm)$)   [black-move-place,name=dummy] {$d$};
    \node at ($(dummy.west)+(0mm,0mm)$)  [red-move-place,name=p1] {$3$};
  \end{pgfonlayer}

  \node[regionnode,fit=(p1),name=c3]{};

\begin{pgfonlayer}{background}
    \node [fit= (c0) (c1) (c2) (c3),Regionnode,name=r1] {};
  \end{pgfonlayer}

\draw[line width=1pt,->] ($(r1.east)+(2mm,0mm)$) -- ($(r1.east)+(13mm,0mm)$);
  
\begin{pgfonlayer}{foreground}
\node at ($(c3)+(23mm,0mm)$)  [black-move-place,name=dummy] {$d$};
    \node at ($(dummy.west)+(0mm,3mm)$)  [red-move-place,name=p1] {$6$};
    \node at ($(dummy.west)+(0mm,-3mm)$)  [green-move-place,name=p2] {$4$};
  \end{pgfonlayer}

  \node[regionnode,fit=(p1) (p2),name=c1]{};

 \begin{pgfonlayer}{foreground}
    \node at ($(dummy.east)+(3mm,0mm)$)  [black-move-place,name=dummy] {$d$};
    \node at ($(dummy.west)+(0mm,0mm)$) [blue-move-place,name=p1] {$0$};
  \end{pgfonlayer}

  \node[regionnode,fit=(p1) ,name=c2]{};

\begin{pgfonlayer}{background}
    \node [fit=(c1) (c2),Regionnode,name=r1] {};
  \end{pgfonlayer}

  \begin{pgfonlayer}{foreground}
    \node at ($(dummy.east)+(6mm,0mm)$)   [black-move-place,name=dummy] {$d$};
    \node at ($(dummy.west)+(0mm,3mm)$)  [white-move-place,name=p1] {$2$};
    \node at ($(dummy.west)+(0mm,-3mm)$)  [orange-move-place,name=p2] {$3$};
  \end{pgfonlayer}

  \node[regionnode,fit=(p1) (p2),name=c3]{};

  \begin{pgfonlayer}{foreground}
    \node at ($(dummy.east)+(6mm,0mm)$)   [black-move-place,name=dummy] {$d$};
    \node at ($(dummy.west)+(0mm,3mm)$)  [blue-move-place,name=p1] {$1$};
    \node at ($(dummy.west)+(0mm,-3mm)$)  [red-move-place,name=p2] {$5$};
  \end{pgfonlayer}

  \node[regionnode,fit=(p1) (p2),name=c0]{};

 \begin{pgfonlayer}{foreground}
    \node at ($(dummy.east)+(3mm,0mm)$)    [black-move-place,name=dummy] {$d$};
    \node at ($(dummy.west)+(0mm,3mm)$)  [orange-move-place,name=p1] {$2$};
    \node at ($(dummy.west)+(0mm,-3mm)$)  [green-move-place,name=p2] {$\omega$};
  \end{pgfonlayer}

  \node[regionnode,fit=(p1) (p2),name=c1]{};

 \begin{pgfonlayer}{foreground}
    \node at ($(dummy.east)+(3mm,0mm)$)  [black-move-place,name=dummy] {$d$};
    \node at ($(dummy.west)+(0mm,0mm)$) [white-move-place,name=p1] {$4$};
  \end{pgfonlayer}

  \node[regionnode,fit=(p1) ,name=c2]{};

 \begin{pgfonlayer}{foreground}
    \node at ($(dummy.east)+(3mm,0mm)$)   [black-move-place,name=dummy] {$d$};
    \node at ($(dummy.west)+(0mm,0mm)$)  [red-move-place,name=p1] {$3$};
  \end{pgfonlayer}

  \node[regionnode,fit=(p1),name=c3]{};

\begin{pgfonlayer}{background}
    \node [fit=(c0) (c1) (c2) (c3),Regionnode,name=r1] {};
  \end{pgfonlayer}

\end{tikzpicture}

\caption{Type II Transition.}
\label{type:2:fig}
\end{figure}
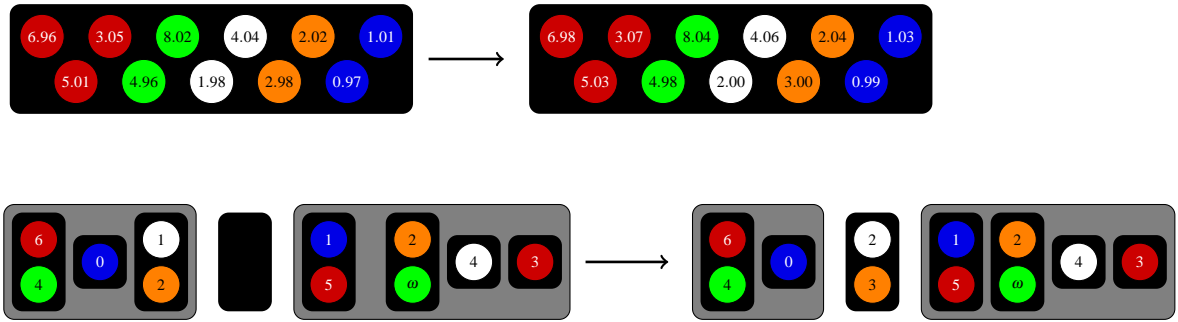

\item[Type III Transition.]
This simulates a delay close to (but smaller than) $1$ 
where the tokens with low fractional parts will now either
have high fractional parts, or
they have reached (and passed) the next integer
and thus have low fractional parts again.
The tokens that already had high fractional parts will all
have passed the next integer and will now have high
fractional parts again.
No token will have an integer value after the transition
(the case where some tokens have integer ages is covered
in Type IV transitions, see below).
Here, the delay is $0.95$.
We have three types of tokens:
\begin{itemize}
\item
Tokens that have low fractional parts both before and after the transition
(the token with age $4.06$ in
\tikz{\node[white-ball]{};}, and the token with age $3.07$ in
\tikz{\node[red-ball]{};}).
The ages of these tokens are 
$5.01$ and $4.02$ after the transition.
Thus, the delay is sufficient to make their ages
go beyond the next integer.
After the transition, these tokens will be  the only ones
with low fractional parts.
The relative ordering of their fractional parts will not be changed.
The integer part of their ages will have increased by one.
At the region level, these two tokens are represented by
the two right-most multisets in $L$.
After the transition, they will be the only multisets in $L$, and
their values are incremented by $1$ each.
Notice that the relative
ordering of these tokens inside the region will be preserved.
\item
Tokens that have low fractional parts before the transition and high
fractional parts after the transition
(the token with age $1.03$ in
\tikz{\node[blue-ball]{};}, 
the token with age $5.03$ in
\tikz{\node[red-ball]{};},
the token with age $2.04$ in
\tikz{\node[orange-ball]{};}, and
the token with age $8.04$ in
\tikz{\node[green-ball]{};}).
The ages of these tokens are 
$1.98$, $5.98$, $2.99$, resp.\ $8.99$ after the transition.
These tokens have the highest fractional parts among all tokens
in the marking.
The relative ordering of the fractional parts of these tokens
will not be changed.
Also, the delay is not  sufficiently long to make their values
reach (or pass) to the next integer.
At the region level, the corresponding multisets
move from $L$ to $H$, and will now be the right-most
multisets in $H$.
The ordering of these multisets is preserved.
\item
Tokens that have high fractional parts both before and after
the transition
(the token with age $6.98$ in
\tikz{\node[red-ball]{};}, 
the token with age $4.98$ in
\tikz{\node[green-ball]{};}, and
the token with age $0.99$ in
\tikz{\node[blue-ball]{};}).
The ages of these tokens are 
$7.93$, $5.93$, resp.\ $1.94$ after the transition.
The delay is  sufficiently long both to make their values
pass the next integer integer, and to make their fractional parts high again.
However, these tokens have now the lowest fractional parts among all tokens
with high fractional parts.
The relative ordering of the fractional parts of the tokens
will not be changed.
At the region level, the corresponding multisets
 will be the left-most
multisets in $H$.
The ordering of these multisets is preserved.
Their values are incremented by one (to reflect that they have
reached the next integer).
Notice that the new value of the token in 
\tikz{\node[red-ball]{};} 
is represented by $\omega$ since
the value is $\geq\maxval+1$.
\end{itemize}

\begin{figure}

\begin{tikzpicture}[]
  \node[name=dummy] {};
  \node at (dummy) [red-move-place,name=n1] {$6.98$};
  \node at ($(n1.east)+(3mm,0mm)$) [red-move-place,name=n3] {$3.07$};
  \node at ($(n3.east)+(3mm,0mm)$) [green-move-place,name=n5] {$8.04$};
  \node at ($(n5.east)+(3mm,0mm)$) [white-move-place,name=n7] {$4.06$};
  \node at ($(n7.east)+(3mm,0mm)$) [orange-move-place,name=n9] {$2.04$};
  \node at ($(n9.east)+(3mm,0mm)$) [blue-move-place,name=n11] {$1.03$};
  \node at ($(n1.center) !.5! (n3.center)+(0mm,-6mm)$) [red-move-place,anchor=center,name=n2] {$5.03$};
  \node at ($(n2.east)+(3mm,0mm)$) [green-move-place,name=n4] {$4.98$};
  \node at ($(n4.east)+(3mm,0mm)$) [white-move-place,name=n6] {$2.00$};
  \node at ($(n6.east)+(3mm,0mm)$) [orange-move-place,name=n8] {$3.00$};
  \node at ($(n8.east)+(3mm,0mm)$) [blue-move-place,name=n10] {$0.99$};
\begin{pgfonlayer}{background}
\node[black-bg,fit=(n1) (n2) (n3) (n4) (n5) (n6) (n7) (n8) (n9) (n10) (n11),name=r]{};
\end{pgfonlayer}

\draw[line width=1pt,->] ($(r.east)+(2mm,0mm)$) -- ($(r.east)+(12mm,0mm)$);

  \node at ($(n11.east)+(18mm,0mm)$)  [red-move-place,name=n1] {$7.93$};
  \node at ($(n1.east)+(3mm,0mm)$) [red-move-place,name=n3] {$4.02$};
  \node at ($(n3.east)+(3mm,0mm)$) [green-move-place,name=n5] {$8.99$};
  \node at ($(n5.east)+(3mm,0mm)$) [white-move-place,name=n7] {$5.01$};
  \node at ($(n7.east)+(3mm,0mm)$) [orange-move-place,name=n9] {$2.99$};
  \node at ($(n9.east)+(3mm,0mm)$) [blue-move-place,name=n11] {$1.98$};
  \node at ($(n1.center) !.5! (n3.center)+(0mm,-6mm)$) [red-move-place,anchor=center,name=n2] {$5.98$};
  \node at ($(n2.east)+(3mm,0mm)$) [green-move-place,name=n4] {$5.93$};
  \node at ($(n4.east)+(3mm,0mm)$) [white-move-place,name=n6] {$2.95$};
  \node at ($(n6.east)+(3mm,0mm)$) [orange-move-place,name=n8] {$3.95$};
  \node at ($(n8.east)+(3mm,0mm)$) [blue-move-place,name=n10] {$1.04$};
\begin{pgfonlayer}{background}
\node[black-bg,fit=(n1) (n2) (n3) (n4) (n5) (n6) (n7) (n8) (n9) (n10) (n11),name=r]{};
\end{pgfonlayer}

%%%%%%%%%%%%%%%%%%  Regions %%%%%%

 \begin{pgfonlayer}{foreground}
\node at  ($(dummy)+(0mm,-30mm)$)  [black-move-place,name=dummy] {$d$};
    \node at ($(dummy.west)+(0mm,3mm)$)  [red-move-place,name=p1] {$6$};
    \node at ($(dummy.west)+(0mm,-3mm)$)  [green-move-place,name=p2] {$4$};
  \end{pgfonlayer}

  \node[regionnode,fit=(p1) (p2),name=c1]{};

 \begin{pgfonlayer}{foreground}
    \node at ($(dummy.east)+(3mm,0mm)$)  [black-move-place,name=dummy] {$d$};
    \node at ($(dummy.west)+(0mm,0mm)$) [blue-move-place,name=p1] {$0$};
  \end{pgfonlayer}

  \node[regionnode,fit=(p1) ,name=c2]{};

\begin{pgfonlayer}{background}
    \node [fit=(c1) (c2),Regionnode,name=r1] {};
  \end{pgfonlayer}

  \begin{pgfonlayer}{foreground}
    \node at ($(dummy.east)+(6mm,0mm)$)   [black-move-place,name=dummy] {$d$};
    \node at ($(dummy.west)+(0mm,3mm)$)  [white-move-place,name=p1] {$2$};
    \node at ($(dummy.west)+(0mm,-3mm)$)  [orange-move-place,name=p2] {$3$};
  \end{pgfonlayer}

  \node[regionnode,fit=(p1) (p2),name=c3]{};

  \begin{pgfonlayer}{foreground}
    \node at ($(dummy.east)+(6mm,0mm)$)   [black-move-place,name=dummy] {$d$};
    \node at ($(dummy.west)+(0mm,3mm)$)  [blue-move-place,name=p1] {$1$};
    \node at ($(dummy.west)+(0mm,-3mm)$)  [red-move-place,name=p2] {$5$};
  \end{pgfonlayer}

  \node[regionnode,fit=(p1) (p2),name=c0]{};

 \begin{pgfonlayer}{foreground}
    \node at ($(dummy.east)+(3mm,0mm)$)    [black-move-place,name=dummy] {$d$};
    \node at ($(dummy.west)+(0mm,3mm)$)  [orange-move-place,name=p1] {$2$};
    \node at ($(dummy.west)+(0mm,-3mm)$)  [green-move-place,name=p2] {$\omega$};
  \end{pgfonlayer}

  \node[regionnode,fit=(p1) (p2),name=c1]{};

 \begin{pgfonlayer}{foreground}
    \node at ($(dummy.east)+(3mm,0mm)$)  [black-move-place,name=dummy] {$d$};
    \node at ($(dummy.west)+(0mm,0mm)$) [white-move-place,name=p1] {$4$};
  \end{pgfonlayer}

  \node[regionnode,fit=(p1) ,name=c2]{};

 \begin{pgfonlayer}{foreground}
    \node at ($(dummy.east)+(3mm,0mm)$)   [black-move-place,name=dummy] {$d$};
    \node at ($(dummy.west)+(0mm,0mm)$)  [red-move-place,name=p1] {$3$};
  \end{pgfonlayer}

  \node[regionnode,fit=(p1),name=c3]{};

\begin{pgfonlayer}{background}
    \node [fit=(c0) (c1) (c2) (c3),Regionnode,name=r1] {};
  \end{pgfonlayer}

\draw[line width=1pt,->] ($(r1.east)+(2mm,0mm)$) -- ($(r1.east)+(13mm,0mm)$);

\begin{pgfonlayer}{foreground}
\node at ($(c3)+(23mm,0mm)$)  [black-move-place,name=dummy] {$d$};
    \node at ($(dummy.west)+(0mm,3mm)$)  [red-move-place,name=p1] {$\omega$};
    \node at ($(dummy.west)+(0mm,-3mm)$)  [green-move-place,name=p2] {$5$};
  \end{pgfonlayer}

  \node[regionnode,fit=(p1) (p2),name=c1]{};

 \begin{pgfonlayer}{foreground}
    \node at ($(dummy.east)+(3mm,0mm)$)  [black-move-place,name=dummy] {$d$};
    \node at ($(dummy.west)+(0mm,0mm)$) [blue-move-place,name=p1] {$1$};
  \end{pgfonlayer}

  \node[regionnode,fit=(p1) ,name=c2]{};

\begin{pgfonlayer}{foreground}
    \node at ($(dummy.east)+(3mm,0mm)$)   [black-move-place,name=dummy] {$d$};
    \node at ($(dummy.west)+(0mm,3mm)$)  [white-move-place,name=p1] {$2$};
    \node at ($(dummy.west)+(0mm,-3mm)$)  [orange-move-place,name=p2] {$3$};
  \end{pgfonlayer}

  \node[regionnode,fit=(p1) (p2),name=c3]{};

  \begin{pgfonlayer}{foreground}
    \node at ($(dummy.east)+(6mm,0mm)$)   [black-move-place,name=dummy] {$d$};
    \node at ($(dummy.west)+(0mm,3mm)$)  [blue-move-place,name=p1] {$1$};
    \node at ($(dummy.west)+(0mm,-3mm)$)  [red-move-place,name=p2] {$5$};
  \end{pgfonlayer}

  \node[regionnode,fit=(p1) (p2),name=c4]{};

 \begin{pgfonlayer}{foreground}
    \node at ($(dummy.east)+(6mm,0mm)$)    [black-move-place,name=dummy] {$d$};
    \node at ($(dummy.west)+(0mm,3mm)$)  [orange-move-place,name=p1] {$2$};
    \node at ($(dummy.west)+(0mm,-3mm)$)  [green-move-place,name=p2] {$\omega$};
  \end{pgfonlayer}

  \node[regionnode,fit=(p1) (p2),name=c5]{};

\begin{pgfonlayer}{background}
    \node [fit=(c1) (c2) (c3) (c4) (c5),Regionnode,name=r1] {};
  \end{pgfonlayer}

\begin{pgfonlayer}{foreground}
    \node at ($(dummy.east)+(6mm,0mm)$)   [black-move-place,name=dummy] {$d$};
    \node at ($(dummy.west)+(0mm,3mm)$)  [black-move-place,name=p1] {$1$};
    \node at ($(dummy.west)+(0mm,-3mm)$)  [black-move-place,name=p2] {$2$};
  \end{pgfonlayer}

  \node[regionnode,fit=(p1) (p2),name=c4]{};

 \begin{pgfonlayer}{foreground}
    \node at ($(dummy.east)+(6mm,0mm)$)  [black-move-place,name=dummy] {$d$};
    \node at ($(dummy.west)+(0mm,0mm)$) [white-move-place,name=p1] {$5$};
  \end{pgfonlayer}

  \node[regionnode,fit=(p1) ,name=c1]{};

 \begin{pgfonlayer}{foreground}
    \node at ($(dummy.east)+(3mm,0mm)$)   [black-move-place,name=dummy] {$d$};
    \node at ($(dummy.west)+(0mm,0mm)$)  [red-move-place,name=p1] {$4$};
  \end{pgfonlayer}

  \node[regionnode,fit=(p1),name=c2]{};

\begin{pgfonlayer}{background}
    \node [fit= (c1) (c2),Regionnode,name=r1] {};
  \end{pgfonlayer}

\end{tikzpicture}

\caption{Type III Transition.}
\label{type:3:fig}
\end{figure}
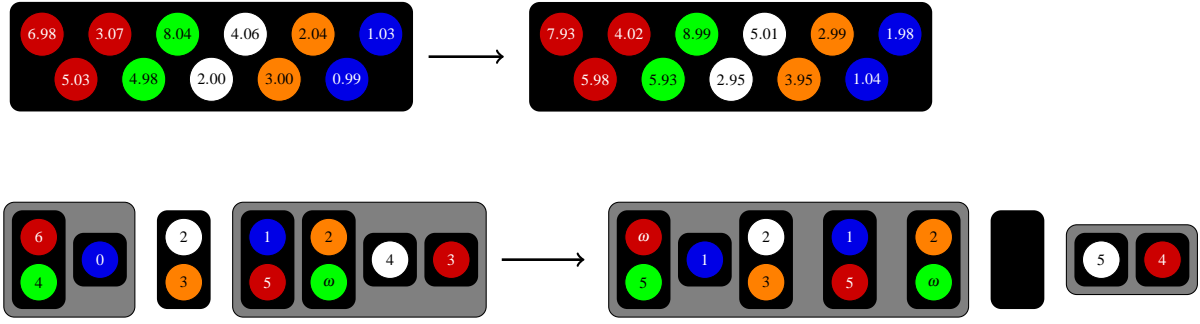

\item[Type IV Transition.]
This is similar to a Type III transition, except that some of the tokens
that have low fractional parts will have integer values after
the transition
(see Figure~\ref{type:4:fig}).

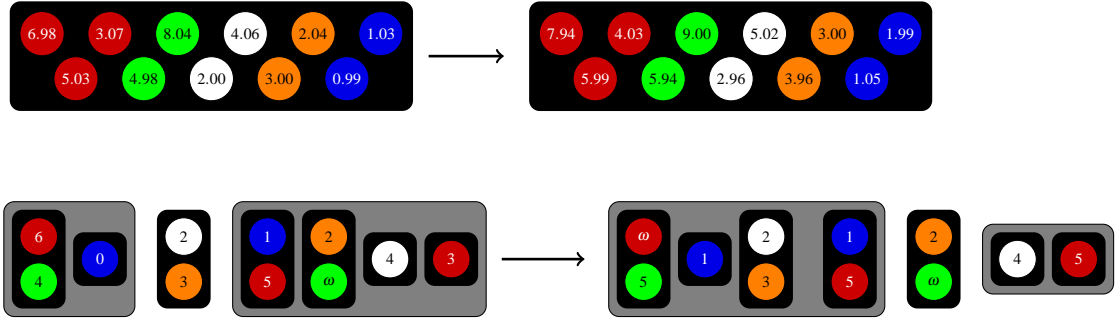
\begin{figure}

\begin{tikzpicture}[]
  \node[name=dummy] {};
  \node at (dummy) [red-move-place,name=n1] {$6.98$};
  \node at ($(n1.east)+(3mm,0mm)$) [red-move-place,name=n3] {$3.07$};
  \node at ($(n3.east)+(3mm,0mm)$) [green-move-place,name=n5] {$8.04$};
  \node at ($(n5.east)+(3mm,0mm)$) [white-move-place,name=n7] {$4.06$};
  \node at ($(n7.east)+(3mm,0mm)$) [orange-move-place,name=n9] {$2.04$};
  \node at ($(n9.east)+(3mm,0mm)$) [blue-move-place,name=n11] {$1.03$};
  \node at ($(n1.center) !.5! (n3.center)+(0mm,-6mm)$) [red-move-place,anchor=center,name=n2] {$5.03$};
  \node at ($(n2.east)+(3mm,0mm)$) [green-move-place,name=n4] {$4.98$};
  \node at ($(n4.east)+(3mm,0mm)$) [white-move-place,name=n6] {$2.00$};
  \node at ($(n6.east)+(3mm,0mm)$) [orange-move-place,name=n8] {$3.00$};
  \node at ($(n8.east)+(3mm,0mm)$) [blue-move-place,name=n10] {$0.99$};
\begin{pgfonlayer}{background}
\node[black-bg,fit=(n1) (n2) (n3) (n4) (n5) (n6) (n7) (n8) (n9) (n10) (n11),name=r]{};
\end{pgfonlayer}

\draw[line width=1pt,->] ($(r.east)+(2mm,0mm)$) -- ($(r.east)+(12mm,0mm)$);

  \node at ($(n11.east)+(18mm,0mm)$)  [red-move-place,name=n1] {$7.94$};
  \node at ($(n1.east)+(3mm,0mm)$) [red-move-place,name=n3] {$4.03$};
  \node at ($(n3.east)+(3mm,0mm)$) [green-move-place,name=n5] {$9.00$};
  \node at ($(n5.east)+(3mm,0mm)$) [white-move-place,name=n7] {$5.02$};
  \node at ($(n7.east)+(3mm,0mm)$) [orange-move-place,name=n9] {$3.00$};
  \node at ($(n9.east)+(3mm,0mm)$) [blue-move-place,name=n11] {$1.99$};
  \node at ($(n1.center) !.5! (n3.center)+(0mm,-6mm)$) [red-move-place,anchor=center,name=n2] {$5.99$};
  \node at ($(n2.east)+(3mm,0mm)$) [green-move-place,name=n4] {$5.94$};
  \node at ($(n4.east)+(3mm,0mm)$) [white-move-place,name=n6] {$2.96$};
  \node at ($(n6.east)+(3mm,0mm)$) [orange-move-place,name=n8] {$3.96$};
  \node at ($(n8.east)+(3mm,0mm)$) [blue-move-place,name=n10] {$1.05$};
\begin{pgfonlayer}{background}
\node[black-bg,fit=(n1) (n2) (n3) (n4) (n5) (n6) (n7) (n8) (n9) (n10) (n11),name=r]{};
\end{pgfonlayer}

%%%%%%%%%%%%%%%%%%  Regions %%%%%%

 \begin{pgfonlayer}{foreground}
\node at  ($(dummy)+(0mm,-30mm)$)  [black-move-place,name=dummy] {$d$};
    \node at ($(dummy.west)+(0mm,3mm)$)  [red-move-place,name=p1] {$6$};
    \node at ($(dummy.west)+(0mm,-3mm)$)  [green-move-place,name=p2] {$4$};
  \end{pgfonlayer}

  \node[regionnode,fit=(p1) (p2),name=c1]{};

 \begin{pgfonlayer}{foreground}
    \node at ($(dummy.east)+(3mm,0mm)$)  [black-move-place,name=dummy] {$d$};
    \node at ($(dummy.west)+(0mm,0mm)$) [blue-move-place,name=p1] {$0$};
  \end{pgfonlayer}

  \node[regionnode,fit=(p1) ,name=c2]{};

\begin{pgfonlayer}{background}
    \node [fit=(c1) (c2),Regionnode,name=r1] {};
  \end{pgfonlayer}

  \begin{pgfonlayer}{foreground}
    \node at ($(dummy.east)+(6mm,0mm)$)   [black-move-place,name=dummy] {$d$};
    \node at ($(dummy.west)+(0mm,3mm)$)  [white-move-place,name=p1] {$2$};
    \node at ($(dummy.west)+(0mm,-3mm)$)  [orange-move-place,name=p2] {$3$};
  \end{pgfonlayer}

  \node[regionnode,fit=(p1) (p2),name=c3]{};

  \begin{pgfonlayer}{foreground}
    \node at ($(dummy.east)+(6mm,0mm)$)   [black-move-place,name=dummy] {$d$};
    \node at ($(dummy.west)+(0mm,3mm)$)  [blue-move-place,name=p1] {$1$};
    \node at ($(dummy.west)+(0mm,-3mm)$)  [red-move-place,name=p2] {$5$};
  \end{pgfonlayer}

  \node[regionnode,fit=(p1) (p2),name=c0]{};

 \begin{pgfonlayer}{foreground}
    \node at ($(dummy.east)+(3mm,0mm)$)    [black-move-place,name=dummy] {$d$};
    \node at ($(dummy.west)+(0mm,3mm)$)  [orange-move-place,name=p1] {$2$};
    \node at ($(dummy.west)+(0mm,-3mm)$)  [green-move-place,name=p2] {$\omega$};
  \end{pgfonlayer}

  \node[regionnode,fit=(p1) (p2),name=c1]{};

 \begin{pgfonlayer}{foreground}
    \node at ($(dummy.east)+(3mm,0mm)$)  [black-move-place,name=dummy] {$d$};
    \node at ($(dummy.west)+(0mm,0mm)$) [white-move-place,name=p1] {$4$};
  \end{pgfonlayer}

  \node[regionnode,fit=(p1) ,name=c2]{};

 \begin{pgfonlayer}{foreground}
    \node at ($(dummy.east)+(3mm,0mm)$)   [black-move-place,name=dummy] {$d$};
    \node at ($(dummy.west)+(0mm,0mm)$)  [red-move-place,name=p1] {$3$};
  \end{pgfonlayer}

  \node[regionnode,fit=(p1),name=c3]{};

\begin{pgfonlayer}{background}
    \node [fit=(c0) (c1) (c2) (c3),Regionnode,name=r1] {};
  \end{pgfonlayer}

\draw[line width=1pt,->] ($(r1.east)+(2mm,0mm)$) -- ($(r1.east)+(13mm,0mm)$);

\begin{pgfonlayer}{foreground}
\node at ($(c3)+(23mm,0mm)$)  [black-move-place,name=dummy] {$d$};
    \node at ($(dummy.west)+(0mm,3mm)$)  [red-move-place,name=p1] {$\omega$};
    \node at ($(dummy.west)+(0mm,-3mm)$)  [green-move-place,name=p2] {$5$};
  \end{pgfonlayer}

  \node[regionnode,fit=(p1) (p2),name=c1]{};

 \begin{pgfonlayer}{foreground}
    \node at ($(dummy.east)+(3mm,0mm)$)  [black-move-place,name=dummy] {$d$};
    \node at ($(dummy.west)+(0mm,0mm)$) [blue-move-place,name=p1] {$1$};
  \end{pgfonlayer}

  \node[regionnode,fit=(p1) ,name=c2]{};

\begin{pgfonlayer}{foreground}
    \node at ($(dummy.east)+(3mm,0mm)$)   [black-move-place,name=dummy] {$d$};
    \node at ($(dummy.west)+(0mm,3mm)$)  [white-move-place,name=p1] {$2$};
    \node at ($(dummy.west)+(0mm,-3mm)$)  [orange-move-place,name=p2] {$3$};
  \end{pgfonlayer}

  \node[regionnode,fit=(p1) (p2),name=c3]{};

  \begin{pgfonlayer}{foreground}
    \node at ($(dummy.east)+(6mm,0mm)$)   [black-move-place,name=dummy] {$d$};
    \node at ($(dummy.west)+(0mm,3mm)$)  [blue-move-place,name=p1] {$1$};
    \node at ($(dummy.west)+(0mm,-3mm)$)  [red-move-place,name=p2] {$5$};
  \end{pgfonlayer}

  \node[regionnode,fit=(p1) (p2),name=c4]{};

\begin{pgfonlayer}{background}
    \node [fit=(c1) (c2) (c3) (c4) ,Regionnode,name=r1] {};
  \end{pgfonlayer}

 \begin{pgfonlayer}{foreground}
    \node at ($(dummy.east)+(6mm,0mm)$)    [black-move-place,name=dummy] {$d$};
    \node at ($(dummy.west)+(0mm,3mm)$)  [orange-move-place,name=p1] {$2$};
    \node at ($(dummy.west)+(0mm,-3mm)$)  [green-move-place,name=p2] {$\omega$};
  \end{pgfonlayer}

  \node[regionnode,fit=(p1) (p2),name=c5]{};

 \begin{pgfonlayer}{foreground}
    \node at ($(dummy.east)+(6mm,0mm)$)  [black-move-place,name=dummy] {$d$};
    \node at ($(dummy.west)+(0mm,0mm)$) [white-move-place,name=p1] {$4$};
  \end{pgfonlayer}

  \node[regionnode,fit=(p1) ,name=c1]{};

 \begin{pgfonlayer}{foreground}
    \node at ($(dummy.east)+(3mm,0mm)$)   [black-move-place,name=dummy] {$d$};
    \node at ($(dummy.west)+(0mm,0mm)$)  [red-move-place,name=p1] {$5$};
  \end{pgfonlayer}

  \node[regionnode,fit=(p1),name=c2]{};

\begin{pgfonlayer}{background}
    \node [fit= (c1) (c2),Regionnode,name=r1] {};
  \end{pgfonlayer}

\end{tikzpicture}

\caption{Type IV Transition.}
\label{type:4:fig}
\end{figure}

\end{description}

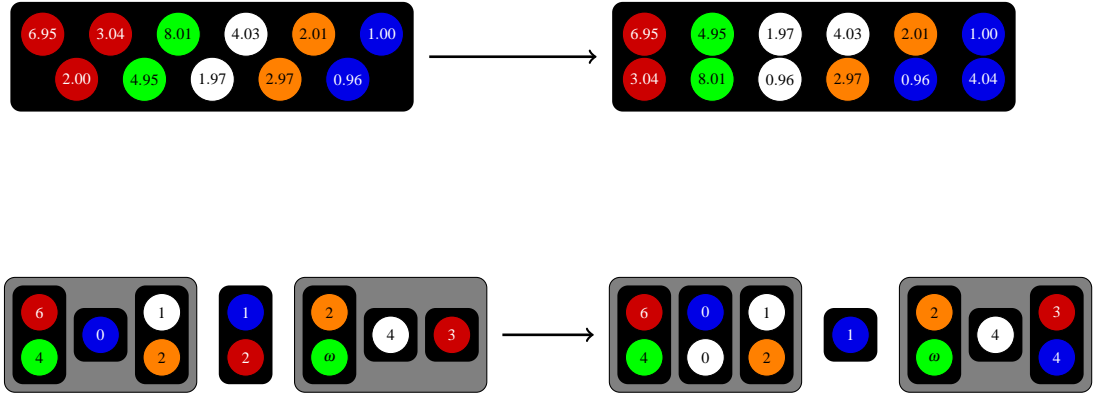
\begin{figure}
\center

\begin{tikzpicture}[]
 \node[name=dummy] {};
 \node[red-move-place,name=n1] at (dummy) {$6.95$};
  \node at ($(n1.east)+(3mm,0mm)$) [red-move-place,name=n3] {$3.04$};
  \node at ($(n3.east)+(3mm,0mm)$) [green-move-place,name=n5] {$8.01$};
  \node at ($(n5.east)+(3mm,0mm)$) [white-move-place,name=n7] {$4.03$};
  \node at ($(n7.east)+(3mm,0mm)$) [orange-move-place,name=n9] {$2.01$};
  \node at ($(n9.east)+(3mm,0mm)$) [blue-move-place,name=n11] {$1.00$};
  \node at ($(n1.center) !.5! (n3.center)+(0mm,-6mm)$) [red-move-place,anchor=center,name=n2] {$2.00$};
  \node at ($(n2.east)+(3mm,0mm)$) [green-move-place,name=n4] {$4.95$};
  \node at ($(n4.east)+(3mm,0mm)$) [white-move-place,name=n6] {$1.97$};
  \node at ($(n6.east)+(3mm,0mm)$) [orange-move-place,name=n8] {$2.97$};
  \node at ($(n8.east)+(3mm,0mm)$) [blue-move-place,name=n10] {$0.96$};
\begin{pgfonlayer}{background}
\node[black-bg,fit=(n1) (n2) (n3) (n4) (n5) (n6) (n7) (n8) (n9) (n10) (n11),name=r1]{};
\end{pgfonlayer}

\begin{scope}[xshift=80mm]
 \node[red-move-place,name=n1] {$6.95$};
  \node at ($(n1.east)+(3mm,0mm)$) [green-move-place,name=n3] {$4.95$};
  \node at ($(n3.east)+(3mm,0mm)$) [white-move-place,name=n5] {$1.97$};
  \node at ($(n5.east)+(3mm,0mm)$) [white-move-place,name=n7] {$4.03$};
  \node at ($(n7.east)+(3mm,0mm)$) [orange-move-place,name=n9] {$2.01$};
  \node at ($(n9.east)+(3mm,0mm)$) [blue-move-place,name=n11] {$1.00$};
  \node at ($(n1.west)+(0mm,-6mm)$) [red-move-place,name=n2] {$3.04$};
  \node at ($(n2.east)+(3mm,0mm)$) [green-move-place,name=n4] {$8.01$};
  \node at ($(n4.east)+(3mm,0mm)$) [white-move-place,name=n6] {$0.96$};
  \node at ($(n6.east)+(3mm,0mm)$) [orange-move-place,name=n8] {$2.97$};
  \node at ($(n8.east)+(3mm,0mm)$) [blue-move-place,name=n10] {$0.96$};
  \node at ($(n10.east)+(3mm,0mm)$) [blue-move-place,name=n12] {$4.04$};
\begin{pgfonlayer}{background}
\node[black-bg,fit=(n1) (n2) (n3) (n4) (n5) (n6) (n7) (n8) (n9) (n10) (n11),name=r2]{};
\end{pgfonlayer}

\end{scope}

\draw[->,line width=1pt] ($(r1.east)+(2mm,0mm)$) -- ($(r2.west)+(-2mm,0mm)$);

 \begin{pgfonlayer}{foreground}
\node at ($(dummy)+(0mm,-40mm)$) [black-move-place,name=dummy] {$d$};
    \node at ($(dummy.west)+(0mm,3mm)$)  [red-move-place,name=p1] {$6$};
    \node at ($(dummy.west)+(0mm,-3mm)$)  [green-move-place,name=p2] {$4$};
  \end{pgfonlayer}

  \node[regionnode,fit=(p1) (p2),name=c1]{};

 \begin{pgfonlayer}{foreground}
    \node at ($(dummy.east)+(3mm,0mm)$)  [black-move-place,name=dummy] {$d$};
    \node at ($(dummy.west)+(0mm,0mm)$) [blue-move-place,name=p1] {$0$};
  \end{pgfonlayer}

  \node[regionnode,fit=(p1) ,name=c2]{};

 \begin{pgfonlayer}{foreground}
    \node at ($(dummy.east)+(3mm,0mm)$)   [black-move-place,name=dummy] {$d$};
    \node at ($(dummy.west)+(0mm,3mm)$)  [white-move-place,name=p1] {$1$};
    \node at ($(dummy.west)+(0mm,-3mm)$)  [orange-move-place,name=p2] {$2$};
  \end{pgfonlayer}

  \node[regionnode,fit=(p1) (p2),name=c3]{};

\begin{pgfonlayer}{background}
    \node [fit=(c1) (c2) (c3),Regionnode,name=r1] {};
  \end{pgfonlayer}

  \begin{pgfonlayer}{foreground}
    \node at ($(dummy.east)+(6mm,0mm)$)   [black-move-place,name=dummy] {$d$};
    \node at ($(dummy.west)+(0mm,3mm)$)  [blue-move-place,name=p1] {$1$};
    \node at ($(dummy.west)+(0mm,-3mm)$)  [red-move-place,name=p2] {$2$};
  \end{pgfonlayer}

  \node[regionnode,fit=(p1) (p2),name=c3]{};

 \begin{pgfonlayer}{foreground}
    \node at ($(dummy.east)+(6mm,0mm)$)    [black-move-place,name=dummy] {$d$};
    \node at ($(dummy.west)+(0mm,3mm)$)  [orange-move-place,name=p1] {$2$};
    \node at ($(dummy.west)+(0mm,-3mm)$)  [green-move-place,name=p2] {$\omega$};
  \end{pgfonlayer}

  \node[regionnode,fit=(p1) (p2),name=c1]{};

 \begin{pgfonlayer}{foreground}
    \node at ($(dummy.east)+(3mm,0mm)$)  [black-move-place,name=dummy] {$d$};
    \node at ($(dummy.west)+(0mm,0mm)$) [white-move-place,name=p1] {$4$};
  \end{pgfonlayer}

  \node[regionnode,fit=(p1) ,name=c2]{};

 \begin{pgfonlayer}{foreground}
    \node at ($(dummy.east)+(3mm,0mm)$)   [black-move-place,name=dummy] {$d$};
    \node at ($(dummy.west)+(0mm,0mm)$)  [red-move-place,name=p1] {$3$};
  \end{pgfonlayer}

  \node[regionnode,fit=(p1),name=c3]{};

\begin{pgfonlayer}{background}
    \node [fit=(c1) (c2) (c3),Regionnode,name=r1] {};
  \end{pgfonlayer}

 \begin{pgfonlayer}{foreground}
    \node at ($(c3)+(23mm,0mm)$) [black-move-place,name=dummy] {$d$};
    \node at ($(dummy.west)+(0mm,3mm)$)  [red-move-place,name=p1] {$6$};
    \node at ($(dummy.west)+(0mm,-3mm)$)  [green-move-place,name=p2] {$4$};
  \end{pgfonlayer}

  \node[regionnode,fit=(p1) (p2),name=c1]{};

 \begin{pgfonlayer}{foreground}
    \node at ($(dummy.east)+(3mm,0mm)$)  [black-move-place,name=dummy] {$d$};
    \node at ($(dummy.west)+(0mm,3mm)$) [blue-move-place,name=p1] {$0$};
    \node at ($(dummy.west)+(0mm,-3mm)$)  [white-move-place,name=p2] {$0$};
  \end{pgfonlayer}

  \node[regionnode,fit=(p1) (p2),name=c2]{};

 \begin{pgfonlayer}{foreground}
    \node at ($(dummy.east)+(3mm,0mm)$)   [black-move-place,name=dummy] {$d$};
    \node at ($(dummy.west)+(0mm,3mm)$)  [white-move-place,name=p1] {$1$};
    \node at ($(dummy.west)+(0mm,-3mm)$)  [orange-move-place,name=p2] {$2$};
  \end{pgfonlayer}

  \node[regionnode,fit=(p1) (p2),name=c3]{};

\begin{pgfonlayer}{background}
    \node [fit=(c1) (c2) (c3),Regionnode,name=r2] {};
  \end{pgfonlayer}

  \begin{pgfonlayer}{foreground}
    \node at ($(dummy.east)+(6mm,0mm)$)   [black-move-place,name=dummy] {$d$};
    \node at ($(dummy.west)+(0mm,0mm)$)  [blue-move-place,name=p1] {$1$};
  \end{pgfonlayer}

  \node[regionnode,fit=(p1),name=c3]{};

 \begin{pgfonlayer}{foreground}
    \node at ($(dummy.east)+(6mm,0mm)$)    [black-move-place,name=dummy] {$d$};
    \node at ($(dummy.west)+(0mm,3mm)$)  [orange-move-place,name=p1] {$2$};
    \node at ($(dummy.west)+(0mm,-3mm)$)  [green-move-place,name=p2] {$\omega$};
  \end{pgfonlayer}

  \node[regionnode,fit=(p1) (p2),name=c1]{};

 \begin{pgfonlayer}{foreground}
    \node at ($(dummy.east)+(3mm,0mm)$)  [black-move-place,name=dummy] {$d$};
    \node at ($(dummy.west)+(0mm,0mm)$) [white-move-place,name=p1] {$4$};
  \end{pgfonlayer}

  \node[regionnode,fit=(p1) ,name=c2]{};

 \begin{pgfonlayer}{foreground}
    \node at ($(dummy.east)+(3mm,0mm)$)   [black-move-place,name=dummy] {$d$};
    \node at ($(dummy.west)+(0mm,3mm)$)  [red-move-place,name=p1] {$3$};
    \node at ($(dummy.west)+(0mm,-3mm)$)  [blue-move-place,name=p2] {$4$};
  \end{pgfonlayer}

  \node[regionnode,fit=(p1) (p2),name=c3]{};

\begin{pgfonlayer}{background}
    \node [fit=(c1) (c2) (c3),Regionnode] {};
  \end{pgfonlayer}

\draw[line width=1pt,->] ($(r1.east)+(2mm,0mm)$) -- ($(r2.west)+(-2mm,0mm)$);  

\end{tikzpicture}
\caption{Firing the transition $t_1$.}
\label{region:disc:fig}
\end{figure}

\paragraph{Discrete Transitions.}
Figure~\ref{region:disc:fig} shows the firing of 
transition $t_1$ (Figure~\ref{ptpn:figure}), and describes how
the firing of the transition may be simulated at the region level.
We remove a token from
\tikz{\node[red-ball]{};}
whose age is in the
interval $[1..3)$.
This is done at the region level by removing  the red ball
with value $2$ from $Z$ (the ball represents a token in
\tikz{\node[red-ball]{};}
whose age is exactly $2$).
We add one to token to
\tikz{\node[white-ball]{};}
whose age is in the interval 
$(0..1)$, and
one to token to
\tikz{\node[blue-ball]{};}
whose age is in the interval 
$[2..5)$.
In Figure~\ref{region:disc:fig}, this is done at the region level by adding 
a white ball to a multiset in $H$
with value $0$ (the ball  represents a token in
\tikz{\node[white-ball]{};}
whose age is in the interval $(0..1)$), and adding 
a blue ball to a multiset in $L$
with value $4$ (the ball  represents a token in
\tikz{\node[blue-ball]{};}
whose age is in the interval $(4..5)$).

\paragraph{Costs.}
At the region level, the cost of performing a type I or type II transition 
is $0$, since we can assume the time delay to be arbitrarily small.
The cost of performing a type III or type IV transition 
is equal to the cost of
performing  a timed transition of $1$ time unit, since we can make
the delay arbitrarily close to $1$.
Thus, the cost of performing the transition in Figure~\ref{type:3:fig} 
or  Figure~\ref{type:4:fig} 
is $15$.
The cost of performing a discrete transition at the region level
is the same as the cost of 
performing the transition on concrete markings.
Thus, the cost of performing the transition in Figure~\ref{region:disc:fig} 
is $2$.

\section{Solving the Cost-Optimality Problem}
\label{solution:section}
In this section we explain our 
solution for the Cost-Optimality problem.
Here, we give an informal overview of the main ideas.
The (quite complicated) technical details 
can be found in \cite{abdulla2011computing}.
First, we show that the 
Cost-Optimality problem can be reduced to the Cost-Threshold
problem.
Then, we introduce a general framework of ordered transition
systems, which we then instantiate to the case of regions.
Finally, we present an algorithm that allows to
solve the Cost-Threshold problem.
\paragraph{From Cost-Optimality to Cost-Threshold.}
Consider an instance  the Cost-Optimality problem,
defined by  $\initmarking$ and $\marking_{\finalp}$ 
(see Section~\ref{ptpn:section}).
The task is to compute the optimal cost of
reaching $\marking_{\finalp}$ from $\initmarking$, i.e.,
the infimum of the costs of all computations
reaching $\marking_{\finalp}$ from $\initmarking$.
To compute this value, it suffices to solve the Cost-Threshold problem 
for any given threshold $\threshold\in\nat$, i.e., to decide whether there
is any computation from  $\initmarking$ to  $\marking_{\finalp}$
with cost $\leq\threshold$.
To see this, we first decide whether $\marking_{\finalp}$
is reachable from $\initmarking$ in the underlying timed Petri net
(without considering costs).
This can be reduced to the Cost-Threshold problem by setting all place and
transition costs to zero and solving the Cost-Threshold problem for
$\threshold=0$. If the answer is no, then we can define the optimal cost
to be $\infty$ ($\marking_{\finalp}$ is not reachable form $\initmarking$).
If yes, then we can find the optimal cost $\threshold$ by solving the
Cost-Threshold problem for threshold $\threshold=0,1,2,3,\dots$ 
until the answer is yes.
We solve the Cost-Threshold problem using regions
as symbolic encodings of sets of markings.
\paragraph{Ordered Transition Systems.}
An {\it ordered transition system}
is a triple $\ts=\tuple{\confset,\amovesto{},\ordering}$
where $\confset$ is a (potentially) infinite set of 
{\it configurations} (or {\it states}),
$\movesto{}$ is a transition relation on $\confset$, and
$\ordering$ is an ordering on $\confset$.
We say that $\movesto{}$ is {\it monotone} wrt.\ $\ordering$ if
the following holds for all configurations
$\conf_1,\conf_2,\conf_3\in\confset$:
if
$\conf_1\movesto{}\conf_2$ and
$\conf_1\ordering\conf_3$ then there is
a $\conf_4$ such that
$\conf_3\movesto{}\conf_4$ and
$\conf_2\ordering\conf_4$.

For a set $\confs\subseteq\confset$ of configurations,
we define $\pre(\confs)$ to be the set of
predecessors of $\confs$ wrt.\ $\movesto{}$, i.e., the set of 
configurations from which we can reach a configuration in $\confs$
through a single application (a single step) of $\movesto{}$.
We define $\pre^*$  to be the reflexive transitive closure
of $\pre$, i.e.,
$\pre^*(\confs)$ is the set of
configurations from which we can reach a configuration in $\confs$
through any  number of steps of $\movesto{}$.

A set $\confs\subseteq\confset$ is said to be
{\it upward-closed} if for any two configurations
with $\conf_1\ordering\conf_2$, it is the case
that $\conf_1\in\confs$ implies $\conf_2\in\confs$.
The {\it upward closure} $\ucof{\confs}$ of a set
$\confs$ of configurations is the set of configurations that
are larger than or equal to some configuration in $\confs$
wrt.\ $\ordering$,
i.e.,
$\ucof{\confs}:=\setcomp{\conf'\in\confset}{\exists\conf\in\confs.\,
\conf\ordering\conf'}$.
Below, we will consider different transition systems that are induced 
by different sets of configurations and different transition relations.

\begin{figure}
\begin{tikzpicture}[]
\begin{scope}[xshift=0mm]

 \begin{pgfonlayer}{foreground}
\node [black-move-place,name=dummy] {$d$};
    \node at ($(dummy.west)+(0mm,3mm)$)  [red-move-place,name=p1] {$6$};
    \node at ($(dummy.west)+(0mm,-3mm)$)  [green-move-place,name=p2] {$4$};
  \end{pgfonlayer}

  \node[regionnode,fit=(p1) (p2),name=c1]{};

 \begin{pgfonlayer}{foreground}
    \node at ($(dummy.east)+(3mm,0mm)$)   [black-move-place,name=dummy] {$d$};
    \node at ($(dummy.west)+(0mm,3mm)$)  [white-move-place,name=p1] {$1$};
    \node at ($(dummy.west)+(0mm,-3mm)$)  [orange-move-place,name=p2] {$2$};
  \end{pgfonlayer}

  \node[regionnode,fit=(p1) (p2),name=c2]{};

\begin{pgfonlayer}{background}
    \node [fit=(c1) (c2),Regionnode] {};
  \end{pgfonlayer}

  \begin{pgfonlayer}{foreground}
    \node at ($(dummy.east)+(6mm,0mm)$)   [black-move-place,name=dummy] {$d$};
    \node at ($(dummy.west)+(0mm,3mm)$)  [blue-move-place,name=p1] {$1$};
    \node at ($(dummy.west)+(0mm,-3mm)$)  [red-move-place,name=p2] {$2$};
  \end{pgfonlayer}

  \node[regionnode,fit=(p1) (p2),name=c3]{};

 \begin{pgfonlayer}{foreground}
    \node at ($(dummy.east)+(6mm,0mm)$)    [black-move-place,name=dummy] {$d$};
    \node at ($(dummy.west)+(0mm,0mm)$)  [green-move-place,name=p1] {$\omega$};
  \end{pgfonlayer}

  \node[regionnode,fit=(p1),name=c1]{};

 \begin{pgfonlayer}{foreground}
    \node at ($(dummy.east)+(3mm,0mm)$)  [black-move-place,name=dummy] {$d$};
    \node at ($(dummy.west)+(0mm,0mm)$) [white-move-place,name=p1] {$4$};
  \end{pgfonlayer}

  \node[regionnode,fit=(p1) ,name=c2]{};

 \begin{pgfonlayer}{foreground}
    \node at ($(dummy.east)+(3mm,0mm)$)   [black-move-place,name=dummy] {$d$};
    \node at ($(dummy.west)+(0mm,0mm)$)  [red-move-place,name=p1] {$3$};
  \end{pgfonlayer}

  \node[regionnode,fit=(p1),name=c3]{};

\begin{pgfonlayer}{background}
    \node [fit=(c1) (c2) (c3),Regionnode,name=R1] {};
  \end{pgfonlayer}

\end{scope}

\begin{scope}[xshift=80mm]

 \begin{pgfonlayer}{foreground}
\node [black-move-place,name=dummy] {$d$};
    \node at ($(dummy.west)+(0mm,3mm)$)  [red-move-place,name=p1] {$6$};
    \node at ($(dummy.west)+(0mm,-3mm)$)  [green-move-place,name=p2] {$4$};
  \end{pgfonlayer}

  \node[regionnode,fit=(p1) (p2),name=c1]{};

 \begin{pgfonlayer}{foreground}
    \node at ($(dummy.east)+(3mm,0mm)$)  [black-move-place,name=dummy] {$d$};
    \node at ($(dummy.west)+(0mm,0mm)$) [blue-move-place,name=p1] {$0$};
  \end{pgfonlayer}

  \node[regionnode,fit=(p1) ,name=c2]{};

 \begin{pgfonlayer}{foreground}
    \node at ($(dummy.east)+(3mm,0mm)$)   [black-move-place,name=dummy] {$d$};
    \node at ($(dummy.west)+(0mm,3mm)$)  [white-move-place,name=p1] {$1$};
    \node at ($(dummy.west)+(0mm,-3mm)$)  [orange-move-place,name=p2] {$2$};
  \end{pgfonlayer}

  \node[regionnode,fit=(p1) (p2),name=c3]{};

\begin{pgfonlayer}{background}
    \node [fit=(c1) (c2) (c3),Regionnode,name=R2] {};
  \end{pgfonlayer}

  \begin{pgfonlayer}{foreground}
    \node at ($(dummy.east)+(6mm,0mm)$)   [black-move-place,name=dummy] {$d$};
    \node at ($(dummy.west)+(0mm,3mm)$)  [blue-move-place,name=p1] {$1$};
    \node at ($(dummy.west)+(0mm,-3mm)$)  [red-move-place,name=p2] {$2$};
  \end{pgfonlayer}

  \node[regionnode,fit=(p1) (p2),name=c3]{};

 \begin{pgfonlayer}{foreground}
    \node at ($(dummy.east)+(6mm,0mm)$)    [black-move-place,name=dummy] {$d$};
    \node at ($(dummy.west)+(0mm,3mm)$)  [orange-move-place,name=p1] {$2$};
    \node at ($(dummy.west)+(0mm,-3mm)$)  [green-move-place,name=p2] {$\omega$};
  \end{pgfonlayer}

  \node[regionnode,fit=(p1) (p2),name=c1]{};

 \begin{pgfonlayer}{foreground}
    \node at ($(dummy.east)+(3mm,0mm)$)  [black-move-place,name=dummy] {$d$};
    \node at ($(dummy.west)+(0mm,0mm)$) [white-move-place,name=p1] {$4$};
  \end{pgfonlayer}

  \node[regionnode,fit=(p1) ,name=c2]{};

 \begin{pgfonlayer}{foreground}
    \node at ($(dummy.east)+(3mm,0mm)$)   [black-move-place,name=dummy] {$d$};
    \node at ($(dummy.west)+(0mm,0mm)$)  [red-move-place,name=p1] {$3$};
  \end{pgfonlayer}

  \node[regionnode,fit=(p1),name=c3]{};

\begin{pgfonlayer}{background}
    \node [fit=(c1) (c2) (c3),Regionnode] {};
  \end{pgfonlayer}

\end{scope}

\node at ($ (R1.east) !.5! (R2.west) $) {\huge$\fordering$};
\end{tikzpicture}
\caption{Ordering on Regions.}
\label{fordering:fig}
\end{figure}
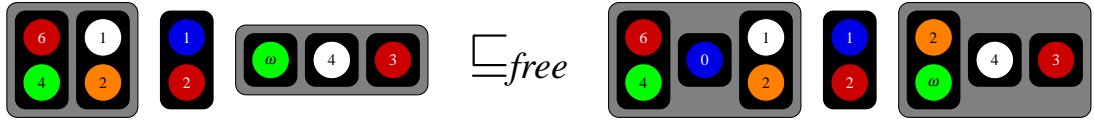

\paragraph{Instantiation.}
Consider an 
instance of the Cost-Threshold problem,
defined by $\initmarking$, $\marking_{\finalp}$, and 
a threshold $\threshold$.
Define a configuration $\conf$ to be a pair
$\tuple{\region,\remainder}$ where $\region$ is a region, and
$\remainder\leq\threshold$. Intuitively, $\remainder$ denotes the maximal
allowed cost of the remainder of a computation that passes through $\region$.
Let $\confset$ be the set of all configurations.
Let $\costconfset$ be the set of configurations
of the form $\tuple{\region,\remainder}$ where
$\region$ contains only tokens in the costs places
(places whose costs are larger than $0$), and where
the number of tokens in $\region$ is smaller
than $\remainder$. 
Notice that 
$\costconfset$  is finite.
Consider regions $\region_1,\region_2$.
We write $\region_1\allordering\region_2$ if we can obtain
$\region_2$ from $\region_1$ by adding a number of tokens
to $\region_1$.
We write $\region_1\fordering\region_2$ if we can obtain
$\region_2$ from $\region_1$ by adding a number of tokens to the free places
(places whose costs are $0$).
Notice that $\fordering\subseteq\allordering$.
Figure~\ref{fordering:fig} shows an example of two regions 
(interpreted over the {\sc Ptpn} of Figure~\ref{ptpn:figure}) related
by $\fordering$.
For configurations 
$\conf_1=\tuple{\region_1,\remainder_1}$
and
$\conf_2=\tuple{\region_2,\remainder_2}$,
we use 
$\conf_1\allordering\conf_2$ 
resp.\
$\conf_1\fordering\conf_2$ 
to denote that
$\remainder_1=\remainder_2$ and that
$\region_1\allordering\region_2$ 
resp.\
$\region_1\fordering\region_2$.
For a set $\confs\subseteq\confset$ of configurations,
we use $\freeucof\confs$ to be the {\it upward closure}
of $\confs$ with respect to $\fordering$, i.e., it contains
all configurations that are larger than or
equal to some configuration in $\confs$ wrt.\ $\fordering$.
We define  $\allucof\confs$ in a similar manner.

Let $\movesto{}_i$ denote the timed transition
relation of type $i\in\set{I,II,III,IV}$, and
let $\discmovesto$ be the discrete transition relation.
Define 
$\movesto{}_A:=\movesto{}_1\cup\movesto{}_2\cup\discmovesto$,
i.e., a transition of type $A$ is either a timed
transition of type I or II, or a discrete transition.
Define 
$\movesto{}_B:=\movesto{}_3\cup\movesto{}_4$,
i.e., a transition of type $B$ is a timed
transition of type III or IV.
For a set $\markings$,
we define $\pre_A(\markings)$ to be the set of
markings from which we can reach a marking in $\markings$
through a single application of a transition of type $A$.
We define $\pre_B(\markings)$ analogously.
%
%We define $\pre_A^*$ and  $\pre_B^*$ to be the reflexive transitive closure of $\pre_A$ resp.\ $\pre_B$.
%

\paragraph{Algorithm.}
%The relation $\amovesto{}$ is monotone wrt.\ $\allordering$, i.e., if $\conf_1\amovesto{}\conf_2$ and $\conf_1\allordering\conf_3$ then there is a $\conf_4$ such that $\conf_3\amovesto{}\conf_4$ and $\conf_2\allordering\conf_4$.
%
We give an overview of an algorithm to solve
the reachability problem.
We notice that $\marking_{\finalp}$ is reachable from 
$\initmarking$ with a cost $\leq\threshold$
iff 
$\initmarking\amovesto{*}\cdot\left(\bmovesto{}\cdot\amovesto{*}\right)^+\marking_{\finalp}$
and the accumulated cost of all involved transitions is  $\leq\threshold$.
Furthermore, we observe that $\marking_{\finalp}$ can be characterized by the upward
closure (wrt.\ $\allordering$) of a finite set of regions.
Therefore, it is sufficient to give an algorithm that,
given a region $\finalregion$ and threshold $\threshold$, checks
whether there is a region $\initregion$ where
$\initmarking$ is included in the denotation of $\initregion$ such that
 $\tuple{\initregion,0}\amovesto{*}\cdot\left(\bmovesto{}\cdot\amovesto{*}\right)^+\allucof{\tuple{\finalregion,\threshold}}$.
To do that, we generate a sequence of sets of configurations 
$V_1,U_1,V_2,U_2,\ldots$,
as follows:
\begin{itemize}
\item
$V_1:=\freeminof{\pre_A^*\left(\allucof{(\finalregion,\threshold)}\right)\cap\left(\freeucof{\costconfset}\right)}$.
This set is possible to compute as follows.
The set $\allucof{\tuple{\finalregion,\threshold}}$ is (obviously) upward-closed
wrt.\ $\allordering$.
The relation $\amovesto{}$ is monotone wrt.\ $\allordering$.
We can then use the backward reachability algorithm
(introduced  in \cite{ACJT96}) for
well quasi-ordered systems 
to compute $\allminof{\pre_A^*\left(\allucof{(\finalregion,\threshold)}\right)}$.
The result follows from the fact that both  
$\allminof{\pre_A^*\left(\allucof{(\finalregion,\threshold)}\right)}$ and
$\costconfset$ are finite.

\item
$U_1:=\freeminof{\pre_B(\freeucof{V_1})}$.
This set can be computed by a straightforward application of 
$\bmovesto{}$ on the elements of $V_1$.
Notice that $U_1\subseteq\freeucof{\costconfset}$, and that it
 is a finite set.

\item
For $k>1$, 
given the finite set $U_k$,
we compute 
$V_k:=\freeminof{\pre_A^*(\freeucof{U_k})\cap\left(\freeucof{\costconfset}\right)}$.
Notice that we here are solving a {\it reachability} 
problem rather than {\it coverability} problem,
since $\freeucof{U_k}$ is not upward-closed wrt.\ $\allordering$.
In fact, this problem has an extremely complicated solution
(described in \cite{abdulla2011computing}).
The construction to compute it uses many calls to a subroutine
which relies on the decidability of the reachability problem for Petri nets 
with one inhibitor arc \cite{Reinhardt:inhibitor,Bonnet-mfcs11}.
In a sense, this is unavoidable, since the reverse reduction also holds.
The reachability problem for Petri nets with one inhibitor arc can be
reduced to the zero-cost coverability problem for {\sc Ptpn}, i.e.,
Cost-Threshold with threshold $0$.

\item
For $k>1$, we compute
$U_k:=\freeminof{\pre_B(\freeucof{V_k})}$
in a similar manner to $U_1$.
Notice that $U_k\subseteq\freeucof{\costconfset}$, and that it
 is a finite set.
\end{itemize}
The sequence $\freeucof{U_1}, \freeucof{U_2}, \dots$ is a monotone-increasing
sequence of upward-closed (wrt.\ $\fordering$)
subsets of $\freeucof{\costconfset}$.
This sequence converges, because $\fordering$ is a well-quasi-ordering
on $\freeucof{\costconfset}$.
Therefore, we get $U_n = U_{n+1}$ for some finite index $n$
and $\freeucof{U_n} = \setcomp{\conf}{\conf (\bmovesto{} \amovesto{*})^* \finalregion}$,
because the
transition $\rightarrow_B$ is only enabled in $\freeucof{\costconfset}$.
Finally,  we compute the (finite) set of configurations,
$\allminof{\pre_A^*(\freeucof{U_n})}$, and check whether
the set contains a configuration of the form $\tuple{\initregion,\remainder}$
such that $\initmarking$ belongs to the denotation of 
$\initregion$.

\section{Conclusions and Future Work}
\label{conclusions:section}
We have given an informal description of a method for computing
the infimum of the costs of placing a token in given place of 
a timed Petri net, starting from a given initial marking.
Interesting directions for future work include augmenting time with
other infinite-state discrete models such as push-down systems and 
asynchronously communicating processes, and to add
other quantitative parameters such as probabilistic behaviors.

\bibliographystyle{eptcs}
\bibliography{base,biblio}

\end{document}